\documentclass{article}
\usepackage{graphicx} 
\usepackage{xcolor} 
\usepackage{verbatim}
\usepackage[utf8]{inputenc}
\usepackage{subfigure}
\usepackage[margin=1in]{geometry} 
\usepackage{booktabs} 
\usepackage{siunitx}  
\usepackage[nodisplayskipstretch]{setspace}
\usepackage{newtxtext,newtxmath} 
\usepackage{enumitem}
\usepackage{url}
\usepackage{amsmath,amsfonts}
\usepackage{algorithm,algpseudocode}
\usepackage{lscape}
\usepackage{caption}
\usepackage[american]{babel}
\usepackage{csquotes}
\usepackage[style=apa,sortcites=true,sorting=nyt,backend=biber,uniquename=false,uniquelist=false]{biblatex}
\DeclareLanguageMapping{american}{american-apa}
\usepackage{xr}  
\usepackage{authblk}

\begin{document}

\title{Cost efficiency of fMRI studies using resting-state vs task-based functional connectivity}

\author[1]{Xinzhi Zhang}
\author[2]{Leslie A Hulvershorn}
\author[3]{Todd Constable}
\author[1]{Yize Zhao}
\author[4]{Selena Wang\footnote{The correspondence should be addressed to Selena Wang, selewang@iu.edu}}
\affil[1]{Department of Biostatistics, Yale School of Public Health, 60 College St, New Haven, CT 06510}
\affil[2]{Department of Psychiatry, Indiana University School of Medicine, Indianapolis, IN, United States}
\affil[3]{Department of Radiology and Biomedical Imaging, Yale School of Medicine, Yale University}
\affil[4]{Department of Biostatistics and Health Data Science, Indiana University School of Medicine, 410 West 10th Street, Suite 3000, Indpls., IN  46202}

\maketitle

\begin{abstract}
We investigate whether and how we can improve the cost efficiency of neuroimaging studies with well-tailored fMRI tasks. The comparative study is conducted using a novel network science-driven Bayesian connectome-based predictive method, which incorporates network theories in model building and substantially improves precision and robustness in imaging biomarker detection. The robustness of the method lays the foundation for identifying predictive power differential across fMRI task conditions if such difference exists. When applied to a clinically heterogeneous transdiagnostic cohort, we found shared and distinct functional fingerprints of neuropsychological outcomes across seven fMRI conditions. For example, emotional N-back memory task was found to be less optimal for negative emotion outcomes, and gradual-onset continuous performance task was found to have stronger links with sensitivity and sociability outcomes than with cognitive control outcomes. Together, our results show that there are unique optimal pairings of task-based fMRI conditions and neuropsychological outcomes that should not be ignored when designing well-powered neuroimaging studies.

\end{abstract}

\doublespace

\section{Introduction}

Recent advances in functional magnetic resonance imaging (fMRI) technologies allow quantification of regional pairwise co-activations across the brain, known as functional connectome (FC).   Association-based analyses allow identification of brain regions with significant links with individual outcomes \parencite[e.g.,][]{satterthwaite2015common,marek2019identifying}.  Meanwhile, the rise of precision medicine devising patient-tailored treatment plans has encouraged a shift in methods from association analyses to individual-level connectome-based predictive modeling \parencite[CPM,][]{finn2015functional,shen2017using}. These models are used to recognize and select meaningful patterns from functional connectivity that explain differences across individuals for a range of cognitive and behavioral variants and disease outcomes.

A pervasive challenge in connectome-behavior linking is to achieve maximum study predictive power and reproducibility given limited resources, e.g., cost of obtaining fMRI data per scan and person. In many fMRI studies, resting-state data has been the default condition for linking functional connectomes with behaviors perhaps because the resting-state acquisition protocol is easier to replicate and compare across studies. However, literature has shown that cognitive tasks can amplify differences across individuals in connectivity that are relevant for explaining differences in various behavior outcomes \parencite{finn2017can,chen2023bayesian,finn2021time}. Resting-state data is perhaps the worst data to use for building CPMs \parencite{zhao2023task,finn2021time}. Predictive models built from task-based fMRI data can achieve higher prediction accuracy in independent samples than those built from resting-state fMRI data \parencite{rosenberg2016neuromarker,greene2018task, jiang2020task}. The addition of task fMRI to resting fMRI provides additional benefits to predictive power \parencite{elliott2019general,gao2019combining}, suggesting that task fMRI provides independent and pertinent information for predicting individual outcomes. These findings highlight the importance of task fMRI in investigating neural substrates of behaviors.

Tasks performed during fMRI scanning sessions are often designed to stimulate a wide range of cognitive, emotional, and psychological processes that perturb brain circuits in unique ways. It is unlikely that one single task is ideal to differentiate individuals across all dimensions of outcomes. Different types of task fMRI tend to show varying strengths of predictive power for different behaviors. This differential in predictive power means that there is untapped potential to maximize the return for scientific investment with a given fMRI scan time and sample size by tailoring the fMRI task during scanning to achieve the most predictive power for addressing a particular research question in a specific clinical setting. Indeed, researchers have begun to investigate this possibility. For example, in schizophrenia research, \textcite{gur2014neurocognitive} found that emotion recognition tasks yielded higher diagnostic accuracy compared to attention tasks. In the domain of anxiety disorders, \textcite{etkin2007functional} showed that certain tasks are more effective for distinguishing individuals for anxiety disorders related propensities. Regarding depression, \textcite{fitzgerald2008meta} found that more complex cognitive tasks, such as the Emotional N-back (EN-back) memory task, were more effective in distinguishing patients from healthy controls. Additionally, \textcite{kaiser2015large} demonstrated that different reward-processing tasks can reveal distinct neural biomarkers for depression. These studies collectively illustrate how task-specific fMRI data can be used for certain type of outcomes as different behavioral categories may be more indicative of certain conditions under specific tasks \parencite{wolfers2015estimating}. This underscores the necessity of carefully selecting and designing task-based fMRI conditions to optimize their predictive power for specific behaviors under investigation.

To systematic evaluate the cost efficiency of task versus resting-state fMRI studies, we investigate the efficacy of seven task and resting conditions and their associated brain circuits perturbation that may be beneficial to investigate a wide range of behavioral, emotional, and psychological outcomes. Our study uniquely contributes to existing literature (1) by investigating this issue in a clinical heterogeneous cohort, the transdiagnostic dataset \parencite{greene2022brain} spanning healthy individuals and those with one or more psychiatric disorders and (2) by fitting a novel Bayesian generative model called LatentSNA \parencite{wang2023inference} that has been proven to bypass lower statistical power and replicability issue often found with existing methods. The transdiagnostic dataset is beneficial for investigating the unique fit of task-fMRI to behavior outcomes because the study participants have heterogeneous functional connectivity profiles given their diverse clinical diagnoses. Different fMRI tasks may invoke different excitatory patterns with participants with different clinical diagnoses. The heterogeneity will maximize the predictive power differential if such difference exists. 

Our network science driven Bayesian generative model is beneficial for investigating differences in predictive power across resting- and task-fMRIs for a range of individual outcomes because our model bypasses the lack of power found with existing fMRI-based predictive models. To make a reproducibility-focused design for model building, we provide probabilistic support on the uncertainty of the connectome/behavior state and future outcome prediction.  Compared with existing network neuroscience models, our model incorporates network theories \parencite{sweet2024network} in modeling building (versus relying on graph theory metrics for model interpretation), drawing on insights regarding the universality of the communicative structures of real-world networks \parencite{barabasi2013network}. Shared network architectures \parencite{wohlgemuth2012small,barabasi2013network} inform us with a universal set of mathematical and statistical instruments for network generation. The model has been demonstrated to show substantially improved precision and robustness in imaging biomarker detection.

In this paper, we compare the performance of different types of resting- and task-based fMRI in their capability to predict various neuropsychological measures and investigate differences in identified neuroimaging biomarkers. We define a neuroimaging biomarker as a measurable regional indicator that reflects individual differences in neuropsychological outcomes,  disease processes, or abnormal functionalities \parencite{kyle2010biomarkers}. Specifically, these neuroimaging biomarkers are collections of brain regions associated with specific behaviors. In doing so, we seek to uncover the unique neural signatures manifested through different task fMRIs that are linked with specific neuropsychological outcomes. Through this comprehensive approach, we strive to contribute to scientific discussions around the cost efficiency of fMRI studies to achieve maximum predictability and robustness and to highlight the unique benefits of well-designed task fMRI studies for investigating complex neuropsychological outcomes.

Our results show that certain task fMRI conditions are more optimal for investigating certain types of neuropsychological outcomes than others. Our results show that there is untapped potential to improve the scientific rigor of current neuroimaging study by designing well-tailored tasks for specific neuropsychological outcomes in addition to employing well-powered methods without incurring a substantial amount of additional cost. More specifically, we found that the gradual-onset continuous performance (gradCPT) task---measuring sustained attention \parencite{rosenberg2013sustaining}--- showed the highest prediction accuracy in independent samples for predicting the Psychological Symptoms Inventory (BriefSymp), Negative Emotional Spectrum (Control) and Sociability and Vitality Scale (Sociability) outcomes. The Emotional N-back (EN-back) task---involving emotional working memory \parencite{levens2008emotion}---showed best predicability for Negative Emotional Spectrum (NegEmo), and Sensory and
Emotional Awareness Scale (Sensitivity). The reading the mind in the eyes (Eyes) task---measuring emotion recognition \parencite{baron1997another} ---best predict Emotional Distress Spectrum (Distress), Empathy Engagement
Scale (Empathy), and Positive Emotional Spectrum (PosEmo). These results suggest that certain fMRI tasks can better predict some type of outcomes than others; this insight is crucial for designing cost effective fMRI studies with limited resources. Researchers can leverage insights from our study to select the most beneficial fMRI task for their study, and thus reduce overall cost and improve their study rigor. In addition, we hope to encourage future research to perform a more comprehensive investigation with other populations, such as the neurodevelopmental Adolescent Brain Cognitive Development (ABCD) study and aging cohorts. 

\section{Material and methods}

\subsection{The Transdiagnostic Dataset}
\begin{figure}[h]
  \centering
  \includegraphics[width=1\textwidth]{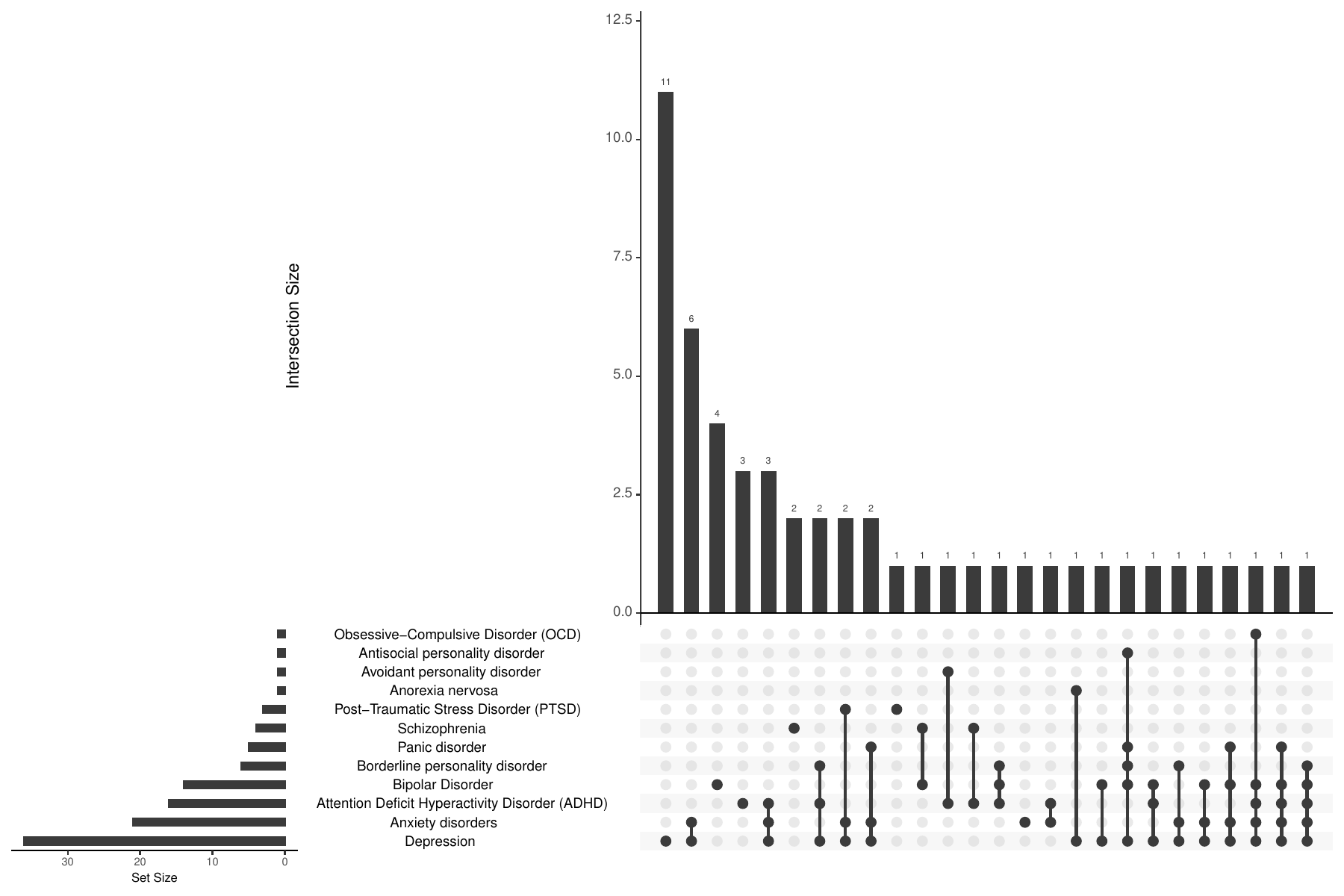}
    \caption{\textbf{Comorbidity patterns among psychiatric disorders} This UpSet plot illustrated our dataset's frequency and co-occurrence of mental health diagnoses. The left axis showed the total number of patients for each disorder, while the main plot displayed the size and composition of various comorbidity combinations. Each column represented a unique combination of disorders, with connected dots below indicating the specific disorders involved. The height of each bar reflected the number of patients with that exact combination of diagnoses.}
  \label{fig:upsetplot}
\end{figure}

We used the transdiagnostic dataset collected at Yale University \parencite{greene2022brain}. We included 190 total participants with 75 females and 54 males. The participants' ages ranged from 19 to 67 years. Among them, 76 were not diagnosed with any mental health condition; 53 had at least one mental health disorder, such as depression, ADHD, psychotic disorder, schizophrenia, or PTSD. Of the 53 participants with psychiatric disorders, 29 had a single diagnosis; 11 patients had 2 diagnoses; 8 patients had 3 diagnoses; 3 had 4 diagnoses; and 2 patients had 5 diagnoses. The co-occurrence of these mental health diseases is shown in Figure~\ref{fig:upsetplot}. The number of patients for each disorder was detailed in Table S1.

With the transdiagnostic dataset, there are seven resting/task conditions for fMRI data including an average condition used as comparison. In addition, participants underwent a structured diagnostic interview and completed a battery of scales derived from validated neuropsychological self-report measures outlined below. Together, with each participant, we have both fMRI data of various conditions and a range of behavior measures.

\subsubsection{Neuropsychological Measures}

As mentioned before, each participant was asked to complete a set of neuropsychological measures, which can be grouped into eight categories, each measuring a distinct latent construct based on the design of the associated survey. We assume unidimensionality of each category of neuropsychological measures though a more comprehensive analysis is warranted to confirm its efficacy and validity; assessing the dimensionality of these surveys is outside of the scope of the current study.  This categorization allows us to perform a more targeted analysis of the relationships between brain activity and neuropsychological outcome. The detailed description of each category is as follows.

\begin{itemize}

    \item \textbf{Negative emotional spectrum (NegEmo)}: The NegEmo category assesses negative emotional experiences. This category focuses on measuring different facets of emotional distress and unpleasant mood states. The NegEmo category is measured by the positive and negative affect schedule - expanded Version \parencite{watson1994panas} including negative affect, fear, sad, guilt, hostility, shyness, and fatigue. 
    \item \textbf{Positive emotional spectrum (PosEmo)}: The PosEmo category measurespleasant emotional states, giving a complete picture of a person's positive feelings and experiences. This category looks at different types of good moods and emotional well-being. It contains six variables: positive affect, joviality, assurance, attentiveness, serenity, and surprise. It is also from positive and negative affect schedule - expanded version \parencite{watson1994panas}. 
    
    \item \textbf{Empathy engagement scale (Empathy)}: The Empathy category evaluates a person's ability to understand and feel other people's emotions. This category examines the different ways that individuals relate to and demonstrate empathy. It contained four variables: perspective taking,
    fantasy, empathic concern, and personal distress. This category comes from the interpersonal reactivity index \parencite{davis1983measuring}.
    
    \item \textbf{Emotional distress spectrum (Distress)}: The Distress category measures emotional distress and negative affective states in adults. This category captures how people experience and express psychological discomfort and unpleasant emotions in response to challenging situations. It contains four measures: fear, sadness, discomfort, and frustration. It is derived from the adult temperament questionnaire \parencite{evans2007developing}.
    
    \item \textbf{Sociability and vitality scale (Sociability)}: The Sociability category captures the tendency to seek out and enjoy social interactions, as well as the experience of high-energy positive emotions. It includes three key variables derived from the surgency part of the adult temperament questionnaire (ATQ) \parencite{evans2007developing}: sociability, positive affect, and high-intensity pleasure.
    
    \item \textbf{Self-regulation control measures (Control)}: The Control category assesses various aspects of an individual's ability to manage their thoughts, emotions, and behaviors. This category captures the capacity for self-regulation and effortful control in adults. It contains three variables from the adult temperament questionnaire \parencite{evans2007developing}: Attentional Control, Inhibitory Control, and Activation Control. 
    
    \item \textbf{Sensory and emotional awareness scale (Sensitivity)}: The Sensitivity category assesses a person's sensitivity to both internal and external stimuli and their level of perceptual awareness. The range and depth of an adult's sensory processing and associative tendencies are summarized in this category. It includes four measures from adult temperament questionnaire \parencite{evans2007developing}: neutral perceptual sensitivity, affective perceptual sensitivity, and associative sensitivity.

    \item \textbf{Psychological symptoms inventory (BriefSymp)}: The BriefSymp category measures a range of psychological symptoms and psychiatric disorders\parencite{derogatis1983brief}. The BriefSymp category contains nine variables: somatization, obsession-compulsion, interpersonal sensitivity, depression, anxiety, hostility, phobic anxiety, paranoid ideation, and psychoticism. 
\end{itemize}

\subsubsection{Functional Connectivity Data}

In our study, we utilized functional connectivity (FC) data derived from seven different fMRI conditions: two resting-state conditions, four task-based conditions, and one average condition. Participants were asked to relax with their eyes open, not engaging in any specific task during resting state conditions. Participants were asked to perform specific cognitive or emotional tasks that activate targeted brain regions during task conditions \parencite{shah2010functional}. These fMRI conditions allow us to explore differences between fMRI conditions and their predicability of different neuropsychological measures. A detailed description of four task-based fMRI is as follows:

\begin{itemize}

\item \textbf{Emotional N-back task (EN-back)}: The EN-back task condition is designed to measure working memory and emotional processing. Participants were asked to identify if an image was the same or different from the one that appeared earlier \parencite{rosenberg2015predicting,tottenham2009nimstim, conley2018racially, gevins1990effects}. 

\item \textbf{Stop-signal task (SST)}: The SST task engages cognitive systems, specifically working memory. Participants must identify whether an image matched the two images presented earlier \parencite{verbruggen2008stop}.

\item \textbf{Reading the mind in the eyes task (Eyes)}: The Eyes task involves cognitive systems and exercises cognitive control. Participants must identify the direction of an arrow stimulus and withhold their response if the arrow turned blue \parencite{cohen1997another}.

\item \textbf{gradual-onset continuous performance task (gradCPT)}:
 The gradCPT task tapes into cognitive systems emphasizing attention. Participants were asked to press a button when they saw images of cities but were asked to refrain from responding when they saw images of mountains \parencite{rosenberg2013sustaining}.

\end{itemize}

\subsubsection{Data pre-processing}

Our data processing pipeline utilized BioImage Suite \parencite{joshi2011unified}, and incorporated several standard preprocessing steps. These included corrections for slice timing and motion, as well as alignment to the MNI template. We followed the methods described in earlier research by \textcite{greene2018task} and \textcite{horien2019individual} to process the fMRI data into functional connectivity matrices. To define brain regions, we employed the Shen-268 atlas, which divides the brain into 268 distinct Regions of Interest (ROIs), also referred to as nodes. For each participant, we computed functional connectivity metrics across various conditions, including task-based and resting states. This process yielded six 268$\times$268 matrices per subject. Each matrix element represented the Pearson correlation coefficient between a pair of brain nodes. To normalize the distribution of these coefficients, we applied Fisher's z-transformation. Additionally, we categorized the brain nodes into ten functional groups based on their anatomical and functional properties. These groups, as detailed by \textcite{shen2010graph}, are Default Mode, Medial Frontal, Fronto-parietal, Motor, Visual I, Visual II, Visual Association, Limbic, Basal Ganglia, and Cerebellum.

\subsection{Methods}

We used the Bayesian generative LatentSNA model to investigate differences in predicative power across task-fMRIs for a range of individual outcomes. In this section, we will first provide an overview of the LatentSNA model, then we will detail the procedures for calculating the prediction accuracy and the process for identifying neuroimaging biomarkers. To maximize replicability and robustness of the results, we employ independent replications of our model results at each behavior-fMRI condition pairing via random sampling (replication robustness) with alternative neuropsychological measures (independent neuropsychological data robustness) and with alternative fMRI conditions (independent fMRI data robustness).

\subsubsection{Overview of LatentSNA Model}


In a previous study \parencite{wang2023inference}, we have shown that by incorporating network theories in the data generation process and model building, we can substantially improve the predicability of an individual's psychological outcomes using fMRI data. This capability of our method allows us to investigate effective differences between fMRI obtained under different task conditions. Satisfactory prediction accuracy in independent samples indicates a level of robustness of the model and serves as an important basis for the comparative analysis. In addition, we identify the critical brain regions linked with neuropsychological outcomes. Lastly, the LatentSNA model allows us to input multiple behavior indicators to precisely measure the corresponding latent construct.

In this section, we provide an overview of the method although due to space constraints, we cannot list all mathematical details here; interested readers should refer to the methodology paper. The LatentSNA method proposes a generative statistical process assuming that each person's $V$ by $V$ connectivity information can be reduced to be a one-dimensional component of $V$ by $1$ dimension. Region-specific connectivity information is then linked with neuropsychological measures to capture significant associations. Different from scalar-on-network regressions \parencite[e.g.,][]{zhao2022bayesian,wang2021learning}, our method proposes a shared data generation process between fMRI and multivariate neuropsychological measures. 
Suppose we have $V\times V$ functional connectivity matrix \( \mathbf{C}_j \) as well as the $P \times 1$ behavioral vector data \( \mathbf{b}_j \) for subject \( j \); $P$ is the total number of variables measuring the unidimensional latent neuropsychological construct. The data generation process for $\mathbf{C}_j$ can be written as $\mathbf{C}_j = \mathbf{D}_j + \mathbf{y}_j \mathbf{y}_j^\top + \mathbf{F}_j$, where $\mathbf{D}_j$ is the intercept matrix, \( \mathbf{y}_j \) represents the latent variable vector of \( \mathbf{C}_j \), and \( \mathbf{F}_j \) is the error term matrix. The data generation process for $\mathbf{b}_j$ can be written as $\mathbf{b}_j = \mathbf{e}_j + \boldsymbol{\kappa}_j + \mathbf{v}_j$, where  \( \mathbf{e}_j \) is the intercept vector, \( \mathbf{k}_j \) represents the latent variable of \( \mathbf{b}_j \), and \( \mathbf{v}_j \) is the error vector term for behaviour. The connectivity and behavioral measures were linked by modeling their latent variables with a shared distribution, where the estimation of these latent variables mutually informs each other.

The model was estimated with a novel MCMC algorithm detailed in \textcite{wang2023inference}.
When fitting the LatentSNA model, we used 5000 burn-in iterations and 15000 MCMC samples. Our model, which incorporated one behavioral variable and functional connectivity data from a fMRI condition for 190 subjects, took approximately 10 hours to fit using a single-core processor and about 8 GB of memory. This timeframe and resource usage represented the computational requirements for fitting just one instance of our model. To avoid being stuck in the local optima, we fitted each model with 10 random parameter initialization and chose the most optimal model based on model fit in the test sample. The process was repeated 10 times for cross-validation. To ensure fair comparison with competing methods, we used identical training and test samples across all methods. The trace plot of the covariance estimate in Figure S5 demonstrated the convergence of our model.

\subsubsection{Predictive Robustness Analysis}

To investigate differences in predictive power among different fMRI task conditions for a range of neuropsychological measures, we applied the LatentSNA method detailed above to predict each neuropsychological variable using each of the four types of brain functional connectivity. We aim to assess the optimality of each fMRI condition for different neuropsychological categories based on their predicative powers. In the follow, we provide details about the procedures.

\begin{itemize}

    \item  Training Sample Selection. We randomly selected the training sample for the neuropsychological measures, which included 90\% of the participants. The remaining 10\% of the participants were reserved as the test sample. This same training and test sample split was used for all models, including the LatentSNA model and the five competing methods detailed below.

    \item Model Training. We trained the LatentSNA model using the functional connectivity of all participants and neuropsychological measures from the training sample to predict the test sample's neuropsychological scores. The five competing methods were trained using the same input data.

    \item Cross-Validation. We repeated previous two steps  five times to perform 5-fold cross-validation for each connectivity and behavior combination.

    \item  Accuracy Measurement. We calculated the prediction accuracy for each model by calculating the correlation between the estimated behavior scores and the actual behavior scores in the test sample.

    \item  Averaging Accuracy Across Folds. We averaged the prediction accuracy across all folds to obtain a final accuracy measure for each model, allowing for a direct comparison between LatentSNA and the five competing methods.

\end{itemize}

In our analyses, we compared the predication accuracy of our LatentSNA model against five existing methods, CPM \parencite{shen2017using,finn2015functional}, Ridge CPM \parencite{rosenblatt2023connectome}, Tensor Network Factorization Analysis  \parencite[TNFA,][]{zhang2019tensor}, Support Vector Machines \parencite[SVM,][]{hearst1998support}, and Random Forest \parencite[RF,][]{belgiu2016random}. We implemented CPM with NetworkToolbox \parencite{R-NetworkToolbox}; we first expanded the functional connectivity matrices into vectors, then used a 0.001 threshold to identify significant connectivity edges as predictors. We implemented Ridge regression with 'glmnet' R package \parencite{R-glmnet}; we used an L2 penalized linear regression and optimized the penalized parameter to handle high-dimensional data. We implemented SVM with 'e1071' \parencite{R-e1071}, and we employed a linear kernel, selected features with correlations above 0.25 with the response, and tuned the cost parameter. We implemented the RF model with 'ranger' \parencite{R-ranger} via 'caret' \parencite{R-caret}. And we underwent a grid search for the optimal number of variables to split at each node (mtry), the splitting rule for tree size (splitrule), and the minimal node size (min.node.size). The number of decision trees used in the Random Forest algorithm is 500. We optimized each model on the training set and evaluated performance by correlating predicted values with observed values on the test set, allowing a thorough comparison of these techniques. For all methods, we used identical training and test sets to ensure fair comparison. The comparison of the prediction accuracy across all methods is shown in Figure~\ref{fig:prediction_maineffect}.


To directly answer the question which fMRI conditions are optimal for investigating a particular type of neuropsychological measures, we conducted a series of regression analyses using the LatentSNA prediction results and summarised results in Table S2. For each of the eight behavior categories (BriefSymp, NegEmo, PosEmo, Empathy, Distress, Sociability, Control, and Sensitivity), we ran a separate model. For each model, the dependent variable was the prediction accuracy for all indicators within the neuropsychological category. This means we analyzed how accurately we could predict all the indicators in a given category (e.g., all indicators of NegEmo) under different fMRI conditions. The predictors were dummy variables representing different fMRI conditions: Rest2, Average, EN-back, SST, Eyes, and gradCPT. The Rest1 condition served as the reference, hence its absence from the predictor list. The mathematical form of the regression model for each neuropsychological category was:


\begin{equation}
    PA = \beta_0 + \beta_1I(\text{Rest2}) + \beta_2I(\text{Average}) + \beta_3I(\text{EN-back}) + \beta_4I(\text{SST}) + \beta_5I(\text{Eyes}) + \beta_6I(\text{gradCPT}),
\end{equation}where $PA$ represents the prediction accuracy for all indicators in the specific neuropsychological category being analyzed under different fMRI conditions; $I(\text{condition})$ is the dummy variable for each fMRI condition; $\beta_0$ represents the baseline prediction accuracy under the Rest1 condition; and $\beta_1$ to $\beta_6$ represent the change in prediction accuracy for each respective fMRI condition compared to Rest1. This analysis, summarized in Table S2, reveals how different fMRI task conditions influenced our ability to predict behavioral variables across various categories. 

\subsubsection{Neuroimaging Biomarker Detection}

In addition to comparing the predictive power of different task fMRI associated with each neuropsychological measure, we are also interested in the differences in identified neuroimaging biomarkers. To do this, we obtained region-specific covariance parameters between behaviors and brain nodes by fitting the LatentSNA model with each category of neuropsychological measures and each fMRI condition. A high covariance/correlation value indicates a strong association between the neuropsychological measure and the regional functional connectivity data. 
 The covariance estimate for each brain region was derived from the posterior mean. To validate our findings, each model underwent a 5-fold cross-validation using randomly chosen training samples, and the average covariance was calculated as the mean of the covariance estimates from the five folds. This approach allowed each model to determine the involvement level of functional biomarkers in a specific neuropsychological category.
 
 After estimating the covariances between behavior categories and brain nodes, we selected the top 20 brain nodes, comprising 10 nodes with the smallest negative covariance values and 10 nodes with the largest positive covariance values. Using the Shen-268 atlas, we mapped each brain node to the corresponding functional networks. This allows us to investigate the comparable involvement of functional systems in each task fMRI-neuropsychology association analyses.

We then summarised the functional biomarkers of each behavior category. The results of the identified neuroimaging biomarkers associated with each neuropsychological category are shown in Result 3.2, where we adopted a spider plot that illustrates the similarities and differences across various behavior categories using the `radarchart` function from the R package `fmsb` \parencite{nakazawa2013functions}.  This plot compared the functional biomarkers of the eight behavioral categories across different fMRI conditions. Each spider plot represented the functional biomarkers of each behavior category. For each spider plot, the input was the number of functional labels for the top 20
brain nodes under all fMRI conditions.

\section{Results}
\subsection{Similarities and Differences Exist in the Predicability of fMRI Task/Resting Conditions.}

\begin{figure}[h]
  \centering
  \includegraphics[width=1\textwidth]{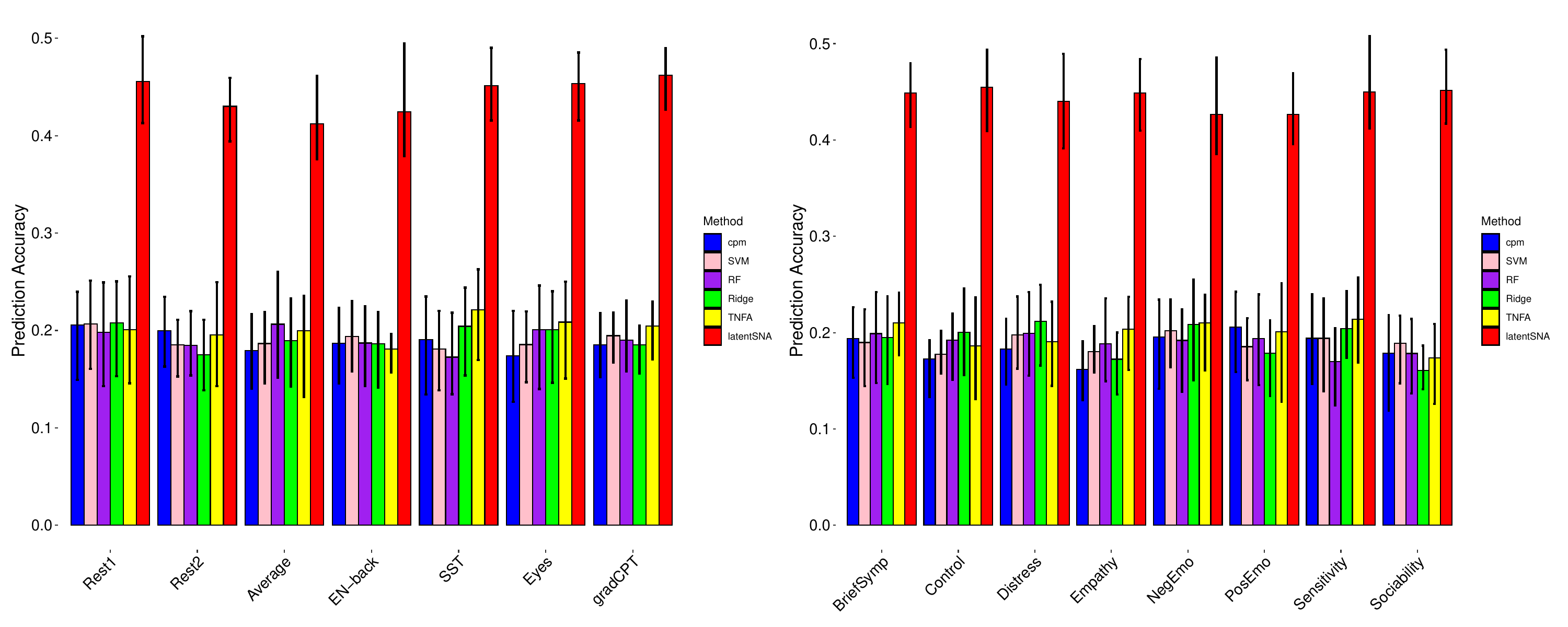}
    \caption {\textbf{Distribution of Prediction Accuracy by Task and behavioral category} The figure presented two histogram plots illustrating the prediction accuracy of latentSNA and 5 competing methods: the left plot by task (Rest1, Eyes, etc.) and the right plot by behavioral category (BriefSymp, Control, etc.). The main bar represented mean prediction accuracy, and the error bar represented 25\% and 75\% quantile, indicating central tendency and variability in prediction across different fMRI conditions and  behavioral categories.}
  \label{fig:prediction_maineffect}
\end{figure}

\begin{figure}[h]
  \centering
  \includegraphics[width=1\textwidth]{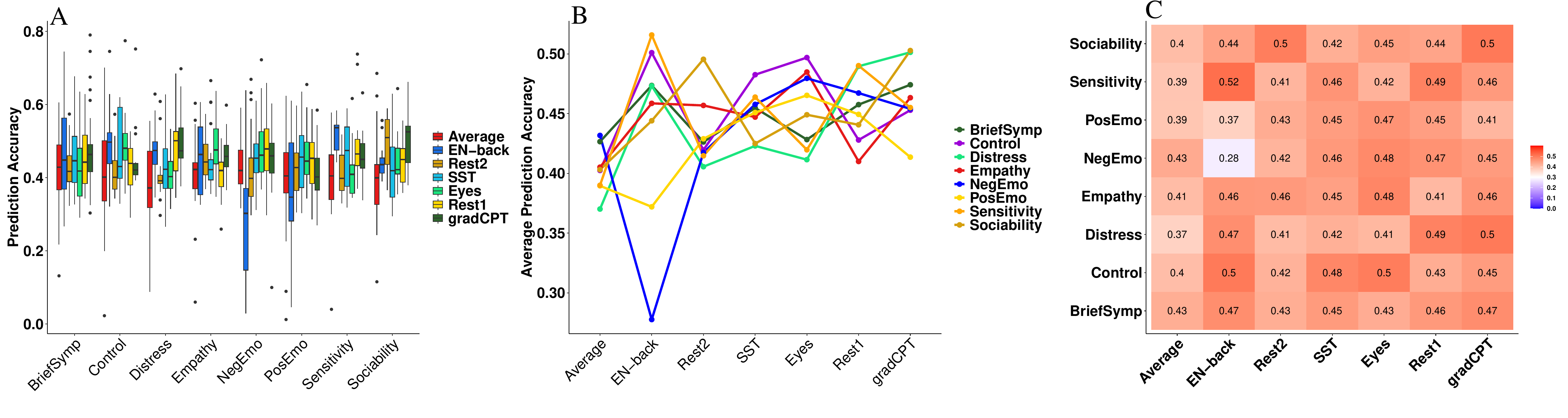}
    \caption{\textbf{Prediction accuracy across tasks and categories} A: Boxplots showed prediction accuracy distributions for different tasks within each category. Colors represented various tasks, illustrating differences in variability and central tendencies. B: Line plot of average prediction accuracy by category across tasks, with each line representing a category. This highlighted performance trends and variations among categories over tasks. C: each cell represented the mean accuracy when predicting a particular behavioral category(y label) using a particular fMRI condition(x label). Calculation of mean prediction accuracy utilized the 5 replications of all behavioral variables in each behavior category.} 
  \label{fig:predict_interacteffect}
\end{figure}

Satisfactory prediction accuracy was found across seven fMRI conditions and eight behavior categories; having a satisfactory fit of the model to data gives us the confidence to further investigate differences in the optimality of fMRI conditions for a range of neuropsychological measures. In Figure~\ref{fig:prediction_maineffect}, we show the prediction accuracy distributions for different tasks within each neuropsychological category (left) as well as for each neuropsychological category across fMRI conditions (right). The average correlations between predicted and observed neuropsychological measures were around 0.45. Interestingly, the Average condition showed lower prediction accuracy than other conditions, suggesting that important features might be lost when averaging across conditions, leading to decreased accuracy and fit across a range of measures. In addition, the EN-back condition exhibited larger standard deviations with a longer error bar, which suggests that compared with other fMRI conditions, the EN-back shows larger differences in its predictability for a range of neuropsychological measures. While some fMRI conditions, such as two resting states show similar predicability across all included outcomes, others such as EN-back shows stronger differences in their predicability for different outcomes. 

We further illustrated the particular pairing of fMRI conditions and neuropsychological measures in Figure~\ref{fig:predict_interacteffect}.  In Figure~\ref{fig:predict_interacteffect}C, we report the average prediction accuracy across 5 replications when predicting a particular neuropsychological category(y label) using a particular fMRI condition(x label). We see that certain tasks were more relevant to some neuropsychological categories and should be prioritized when predicting outcomes in these contexts. 
When predicting outcomes using the EN-back fMRI data,  NegEmo category exhibited substantially lower accuracy than others. The EN-back condition achieved nearly 0.5 prediction accuracy for the Sensitivity category while only 0.27 for the NegEmo category. A further investigation of indicators of the NegEmo category, the average prediction accuracy for fatigue, fear, guilt, hostility, negative affect, sadness, and shyness was 0.35, 0.12, 0.11, 0.18, 0.39, 0.50, 0.33, respectively. The average prediction accuracy was calculated as the mean of five-fold cross-validation. We noticed that EN-back provides high predictions for negative effects, but low prediction accuracy for negative feelings such as fear, guilt, and hostility. The lower prediction accuracy for these negative emotions indicated that the connectome under the EN-back task condition contained limited negative feelings-related features, making EN-back less optimal for investigating negative emotions.

To answer the question---which fMRI condition, resting and task included can best predict a specific category of neuropsychological measures---we coded the fMRI task conditions as predictors of a regression model explaining differences in prediction accuracies (averaged across 5 replications) associated with each category of neuropsychological measures. We report the results in  Table S2. With Rest 1 as the reference condition, a larger coefficient indicated a comparatively higher averaged prediction accuracy under the corresponding fMRI condition. 
With different fMRI conditions as predictors,  the NegEmo category had the largest $R^2$, which suggests there is substantial variation in the predicability of different fMRI conditions for NegEmo measures. We found a large negative coefficient, -0.190, for the EN-back condition, suggesting that EN-back shows substantially lower prediction accuracy in predicting NegEmo than Rest 1. This result corresponds with the result shown in Figure \ref{fig:predict_interacteffect}. Together, our result suggests that similarities and differences exist in the predicability of fMRI task and resting conditions.

\begin{figure}[H]
  \centering
  \includegraphics[width=.8\textwidth]{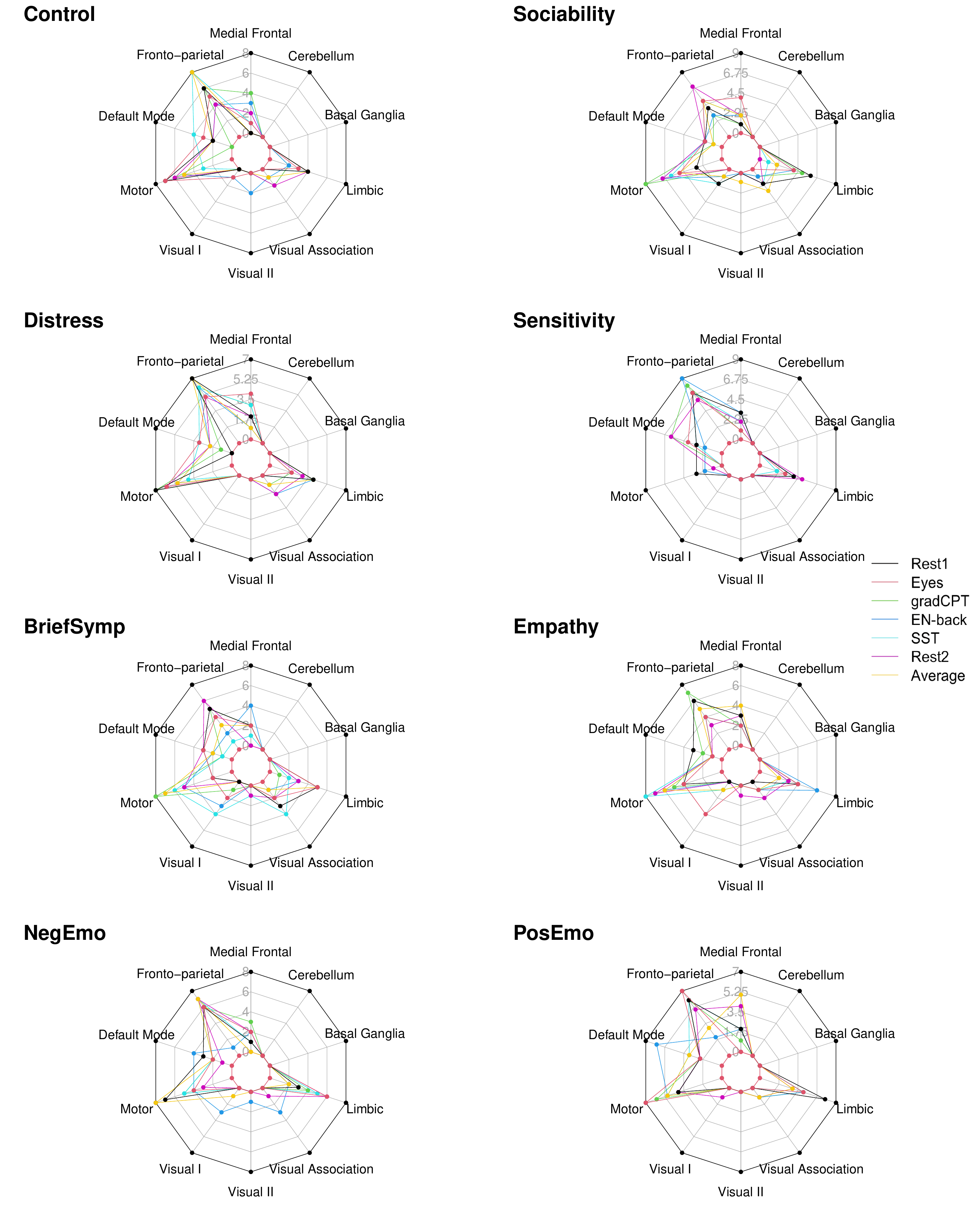}
    \caption {\textbf{Spidar plot of Functional Biomarkers for behavioral categories} This figure displayed spider plots of functional biomarkers across different behavioral categories: Control, Sociability, Distress, Sensitivity, BriefSymp, Empathy, NegEmo, and PosEmo. Each plot represented the distribution of various functional biomarkers in specific anatomical regions, with different colored lines indicating different fMRI conditions. The number of functional biomarkers was derived from the functional labels assigned to the top 20 brain nodes. The top 20 brain nodes referred to the top 10 brain nodes with the largest positive covariance and the top 10 brain nodes with the smallest negative covariance. The covariance estimates between 268 brain nodes and behavior category were calculated by the latentSNA model.}
  \label{fig:spiderplot}
\end{figure}
\subsection{Neuropsychological Measures Show Shared and Distinct Functional Fingerprints}

In Figure~\ref{fig:spiderplot}, we present spider plots for the number of top 20 functional biomarkers in ten functional systems for each category of neuropsychological measure including Control, Sociability, Distress, Sensitivity, BriefSymp, Empathy, NegEmo, and PosEmo.
We use these plots to summarize similarities and differences in patterns of biomarkers across functional systems among neuropsychological categories. Figure~\ref{fig:spiderplot} shows that there are both consistent patterns and notable differences in functional biomarkers across neuropsychological categories. In the following, we provide detailed examination of these patterns to gain insights into the neural underpinnings of various neuropsychological outcomes and identified potential biomarkers for specific neuropsychological categories.

In Figure~\ref{fig:spiderplot}, we consistently identified fronto-parietal, motor, and limbic functional systems to be associated with neuropsychological measures. The fronto-parietal network was found to contribute to individual differences for a range of neuropsychological categories using most fMRI conditions. Our finding aligned with previous research underscoring the integral role of the fronto-parietal network in attention and consciousness \parencite[e.g.,][]{naghavi2005common,hearne2016functional}. Furthermore, \textcite{wang2010effective} emphasized the fronto-parietal network's crucial function in attention control. Building on these existing literature, our results showed that the fronto-parietal network was involved in both the control category and other emotional categories.  Another significant functional network identified was the motor system, which was found to be associated with emotion-related behaviors. This includes facial expression \parencite{morecraft2004motor}, recognition of emotions \parencite{melzer2019we}, and freezing behaviors in response to emotional stimuli \parencite{blakemore2017emotional}. Specifically, the number of brain regions in the motor functional system was comparable to that of the fronto-parietal system associated with the NegEmo, PosEmo, and Distress categories, suggesting that both positive and negative emotions were associated with brain regions in the motor functional system. These findings enhanced our understanding of the role of motor brain regions in emotional processing, building on previous research that had shown their involvement in cognitive and memory-related functions, such as explicit memory \parencite{nyberg2001reactivation} and delayed intentions \parencite{eschen2007motor}. The limbic system was also found to be associated with neuropsychological measures. The limbic system  was associated with NegEmo, PosEmo, Empathy, and Sociability categories, replicating the previous literature supporting the role of the limbic system in emotions \parencite{rolls2015limbic}.

Despite the similarities in the functional fingerprints associated with neuropsychological measures, there were differences as well. It is interesting that the motor functional system was consistently absent in explaining individual differences in Sensitivity across independent replications of the same model and across replications with different fMRI conditions, which contrasted sharply with its role in other neuropsychological categories. The Sensitivity category encompassed both neutral perceptual sensitivity and affective perceptual sensitivity, each tied to perceptual processes. Research established a potential link between motor ability and perceptual development during the infant period \parencite{bushnell1993motor}. Our research suggests that this relationship is mitigated or even lost during adulthood. Additionally, when comparing sensitivity with other neuropsychological measures, the default mode network was found to be dominantly involved in neuropsychological outcomes, especially in the resting state. This observation was consistent with findings in Default Mode—according to \textcite{raichle2015brain}, which showed that the default mode network exhibited increased activity during relaxed, non-task states. 

Furthermore, we noticed that both the fronto-parietal and default mode networks were more associated with Distress compared to Sociability. The Distress category included fear and discomfort, which were related to stress, while Sociability was more associated with pleasure and positive feelings. Our results show that there is consistent involvement of the default mode and fronto-parietal networks in Distress and less so in Sociability, which corresponds to existing literature suggesting that these networks cooperated to produce thoughts \parencite{smallwood2012cooperation} and were more involved in stress and anxiety-related feelings \parencite{dixon2022frontoparietal}.

  
To provide secondary support for the identified shared and distinct functional fingerprints associated with neuropsychological measures, we conducted a robustness check using replication level data---which compared with cross-validated results, shows the model replicability with random sampling of the training and test data. The regression results revealed significant associations that reinforced our previous observations (see details in the Table S3 of the supplementary materials).  From Table S3, we found that brain nodes in the fronto-parietal and motor networks exhibited stronger associations with a range of outcomes. This conclusion was drawn from regressing the absolute values of brain node covariance on network dummy variables, where the coefficients for fronto-parietal and motor networks showed larger magnitudes. These results further reinforced our previous findings that the fronto-parietal and motor networks are involved across categories.

\subsection{Resting and Task fMRI Conditions Reveal Shared and Distinct Brain-behavior Links}
\begin{figure}[H]
  \centering
  \includegraphics[width=0.8\textwidth]{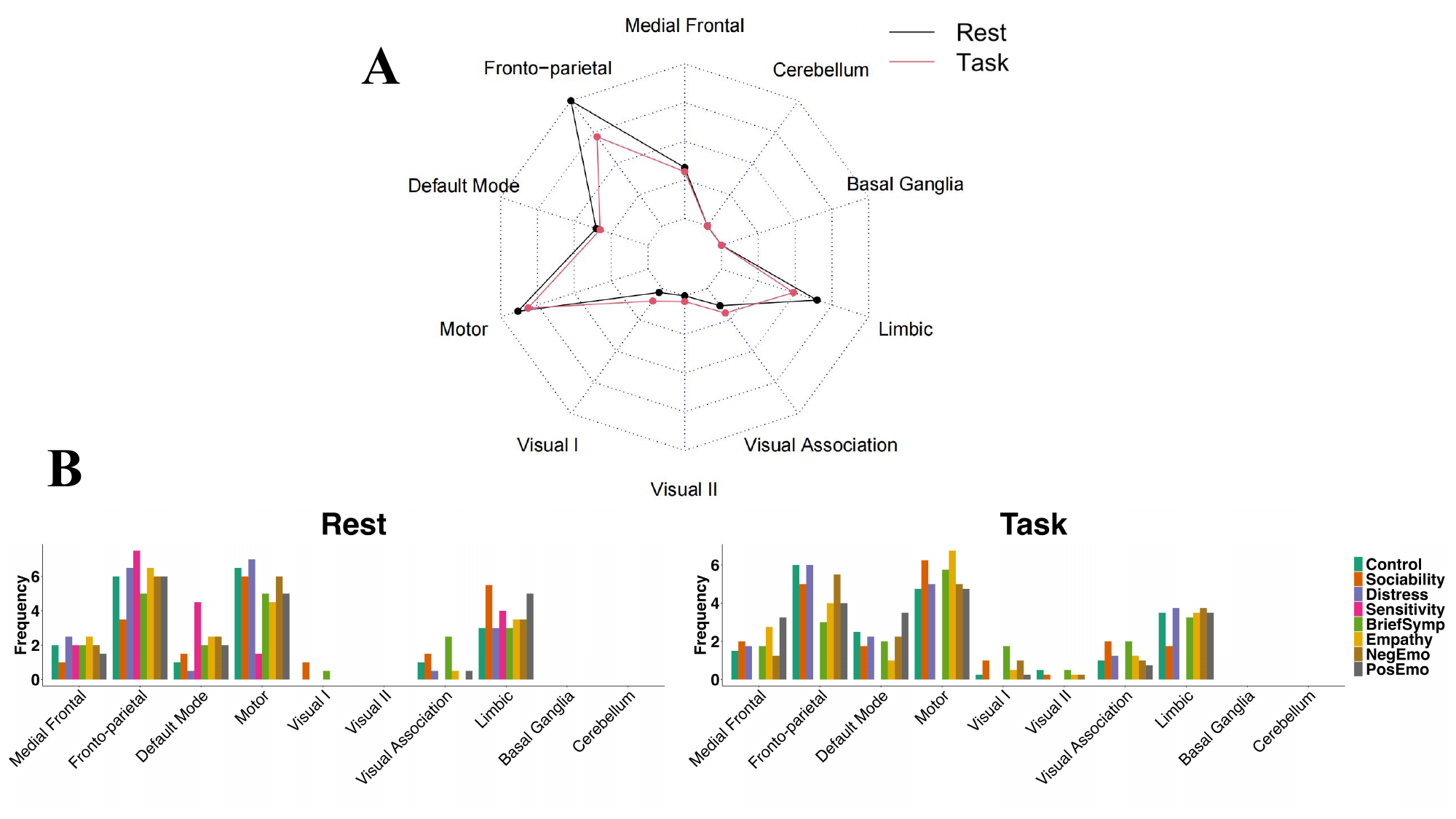}
    \caption {\textbf{Comparison of Functional Biomarkers by Rest and Task Condition} Panel A showed a spider plot comparing the distribution of functional biomarkers between Rest (black line) and Task (red line) conditions across various anatomical regions. Panel B presented bar plots of the frequency of different functional biomarkers in the Rest and Task conditions, with each color representing a different  behavioral category (Control, Sociability, Distress, Sensitivity, BriefSymp, Empathy, NegEmo, PosEmo). To calculate the frequency, we first counted the functional labels assigned to the top 20 brain nodes under each fMRI condition. We then averaged these counts across all task conditions and both resting states to obtain the final frequencies for Rest and Task.}
  \label{fig:rest_task}
\end{figure}

In this section, we perform a comparative analysis of the identified brain-behavior links using resting versus task fMRI as well as between different types of task fMRI conditions. Given that Rest 1 and Rest 2 conditions produced highly comparable results (see Table S1), we combined these fMRI conditions into a single Rest condition for subsequent analyses, enabling a more streamlined comparison of task-based and resting-state functional biomarkers. Tasks performed during fMRI scanning sessions are often designed to stimulate an unique emotional or cognitive response that perturb brain circuits in unique ways. This suggests that there might be differences in activated functional networks corresponding to brain regions engaged during specific tasks conditions.  


We first investigate differences in identified brain-behavior links between resting and task conditions. To do this, we combined two resting states by calculating the mean number of functional biomarkers corresponding to the top 20 brain nodes in Rest 1 and Rest 2. We combined four task conditions by averaging the count of functional biomarkers of the top 20 brain nodes across all task conditions. We compared the frequency of functional biomarker of averaged Task condition with the averaged Rest condition; see results in Figure~\ref{fig:rest_task}. Figure~\ref{fig:rest_task}A showed a spider plot with the count of top 20 brain regions in each of the ten functional systems between rest and task conditions averaged across neuropsychological measures. Figure~\ref{fig:rest_task}B provided bar plots of the frequency of top 20 brain-behavior links in each functional system, for rest and task conditions, broken down by neuropsychological categories. The result showed several key similarities between the Rest and Task conditions. The fronto-parietal biomarkers were prominent in both resting and task conditions, with greater prominence observed during rest. The motor and limbic functional biomarkers were also prevalent with similar frequencies in both resting and task conditions.

\begin{figure}[H]
  \centering
  \includegraphics[width=1\textwidth]{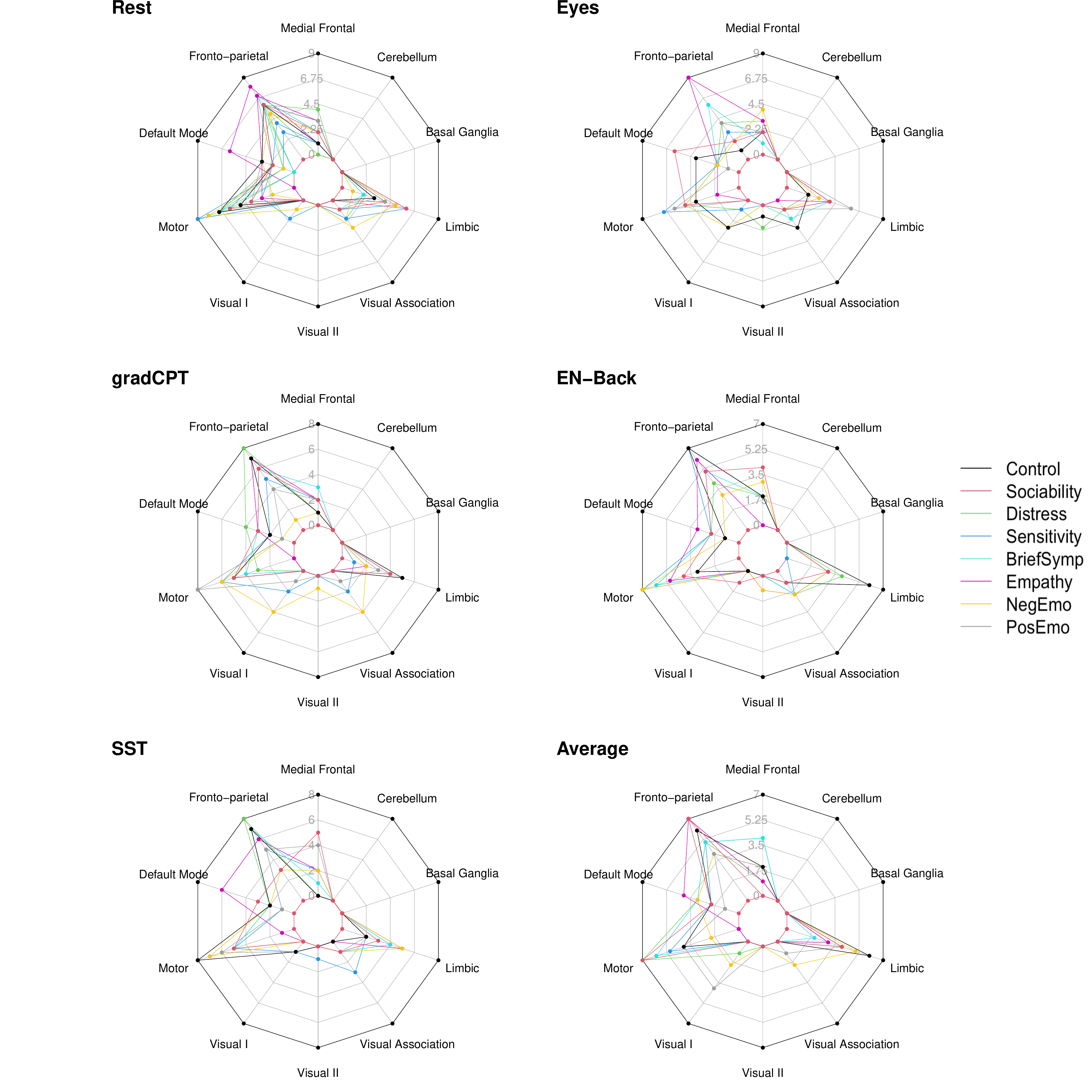}
    \caption {\textbf{Spider plot of functional biomarkers for each fMRI condition} This figure displayed spider plots of functional biomarkers of top 20 brain nodes (The definition of top 20 brain nodes is same as that in Figure~\ref{fig:spiderplot}) across different fMRI conditions: Rest, Eyes, gradCPT, EN-back, SST, and Average. Each plot represented the distribution of various functional biomarkers, with different colored lines indicating different  behavioral categories, such as Control, Sociability, Distress, Sensitivity, BriefSymp, Empathy, NegEmo, and PosEmo. }
  \label{fig:spider_task}
\end{figure}

Having compared rest and task conditions broadly, we then examined differences among specific task conditions. The difference among task conditions was measured by the number of functional biomarkers corresponding to the top 20 brain nodes. The top 20 brain nodes are identified as the combination of the top 10 brain nodes with the highest positive covariance and the top 10 brain nodes with the lowest negative covariance. We first calculate the number of biomarkers for each neuropsychological measure under each fMRI condition using five-fold cross validation. Then, we averaged the number of biomarkers by fMRI conditions and reported results in Table S6. Table S6 detailed that the number of fronto-parietal biomarkers in the Rest condition was comparable to gradCPT, EN-back, and SST, but exceeded that of the Eyes condition. This explained the lower average fronto-parietal biomarker count in task conditions observed in Figure~\ref{fig:rest_task}, driven by the Eyes condition. Figure~\ref{fig:spider_task}  displayed spider plots of functional biomarkers of the top 20 brain nodes across different fMRI conditions: Rest, Eyes, gradCPT, EN-back, SST, and Average.
Figure \ref{fig:spider_task} shows that the Eyes fMRI condition displays a different brain-behavior link pattern than other task conditions. While gradCPT, EN-back, and SST showed predominantly fronto-parietal and motor biomarkers, the Eyes condition additionally displayed default mode and medial frontal involvement. The average condition appeared to occupy an intermediate position between rest and task conditions. More specifically, the Average condition has a similar number of fronto-parietal functional biomarkers as the Rest condition. However, in terms of visual biomarkers, it is closer to that of the gradCPT and Eyes task-based conditions than the Resting State.

\begin{figure}[H]
  \centering
  \includegraphics[width=1\textwidth]{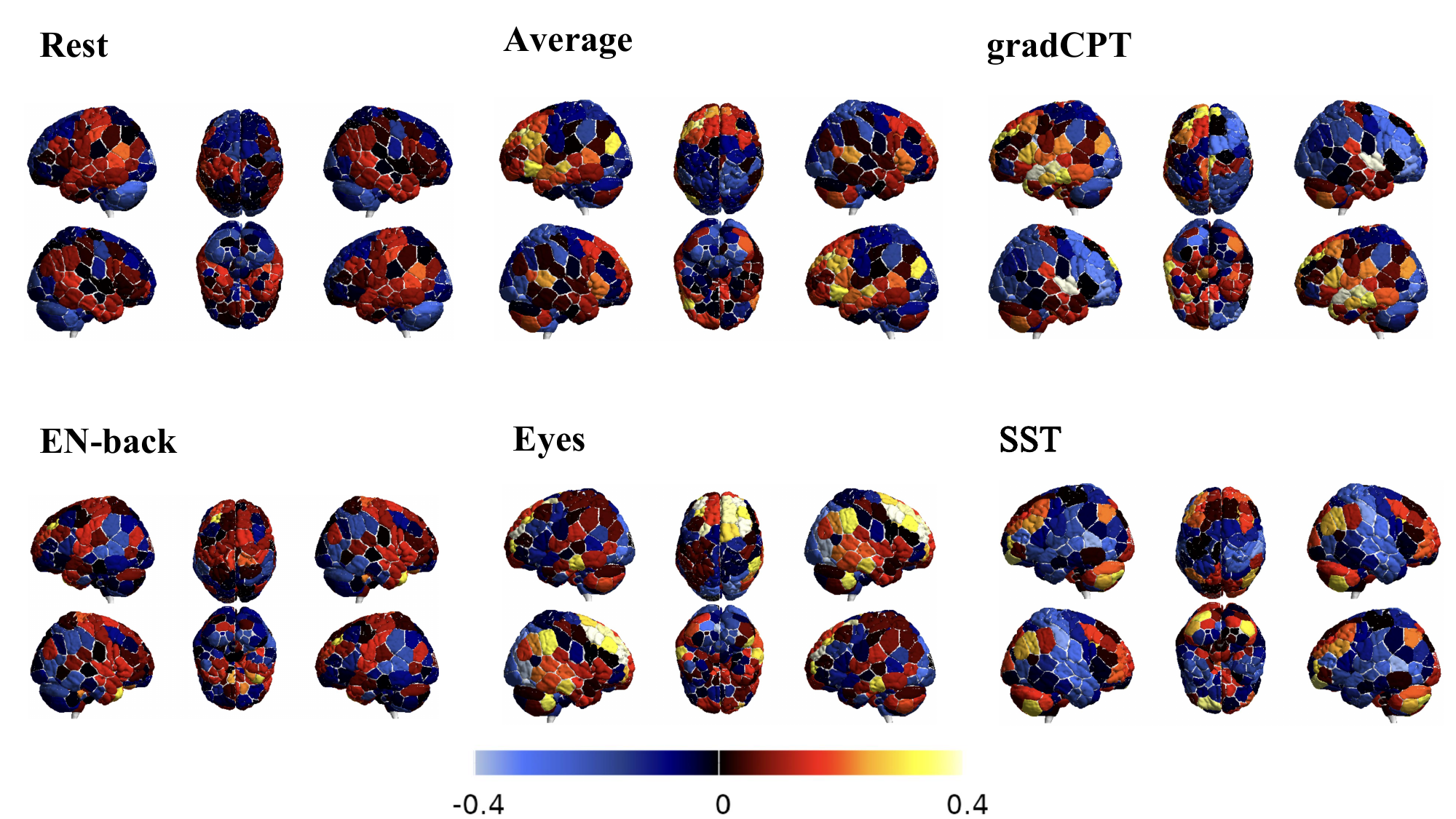}
    \caption {\textbf{Covariance Estimates Between Brain Regions and Sociability Across Various fMRI Conditions} This figure visualized the covariance estimates of 268 brain nodes across different fMRI conditions: Rest, Average, gradCPT, EN-back, Eyes, and SST. The color scale indicated the magnitude of covariance, ranging from -0.4 (blue) to 0.4 (yellow). A lighter color indicated a higher absolute value of covariance. The plot was drawn using the ROI drawing function in BrainNetViewer}
  \label{fig:brainplot}
\end{figure}

\begin{figure}[H]
  \centering
  \includegraphics[width=1\textwidth]{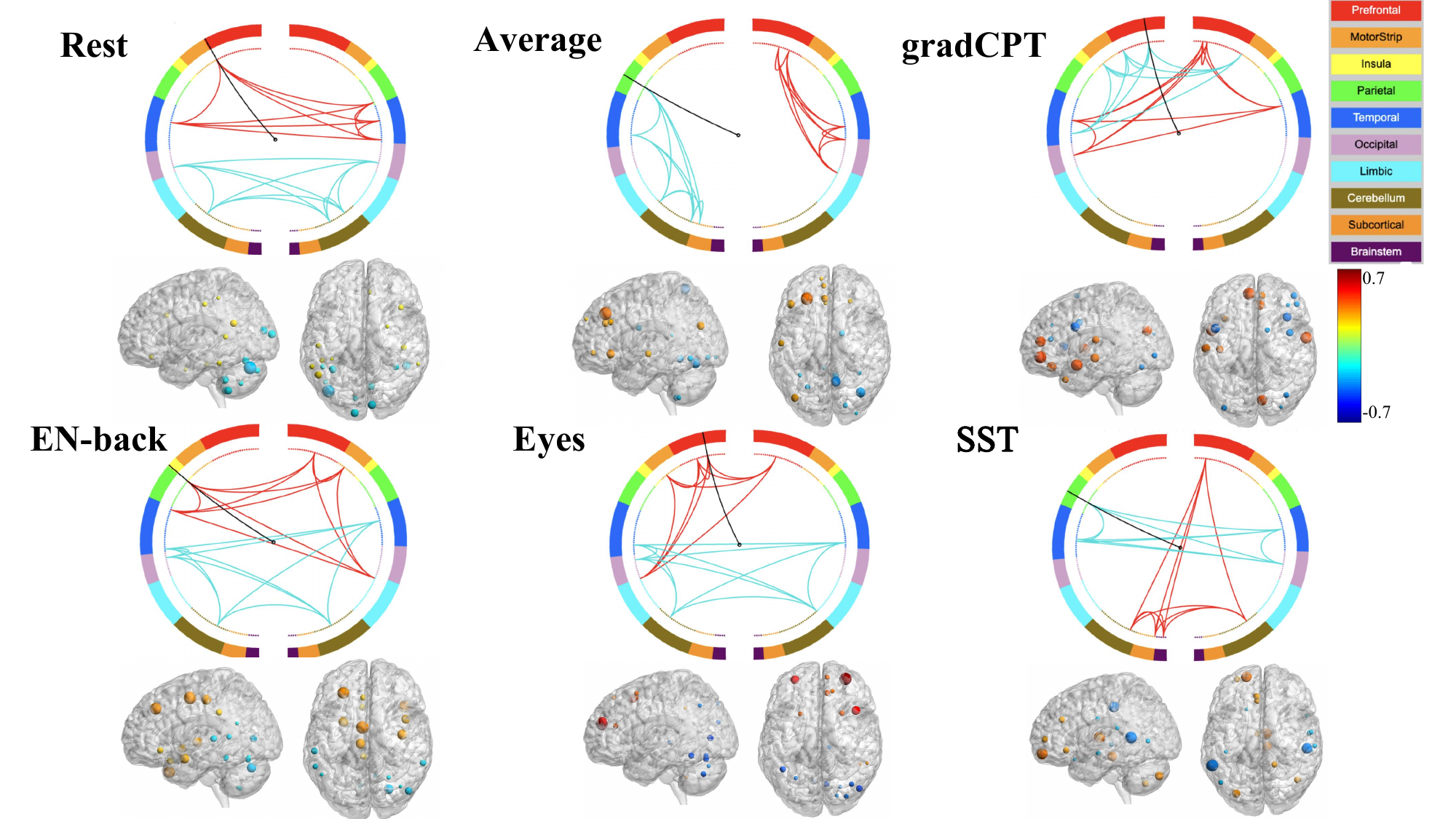}
    \caption {\textbf{Circle plot and brain plot corresponding to top brain nodes for each task condition} The Circle plot shows the connectivity of brain nodes with top 5 positive covariance(in red) and top 5 negative covariance(in blue) for Sociability in each fMRI condition. The brain plot shows the positions of the top 20 brain nodes. The node size reflects the absolute value of the covariance estimate. The color of nodes reflects the covariance, with blue representing negative covariance and red representing positive covariance}
  \label{fig:circle plot}
\end{figure}

We further investigated the covariance estimates for each of the 268 brain regions, representing the strength of their associations with Sociability category in Figure~\ref{fig:brainplot}. We focused on Sociability for further investigation because, on average, the Sociability category resulted in the highest absolute value of covariance estimates of brain nodes under all fMRI conditions. Besides the Socioability measure, we also provide a comprehensive analysis with additional neuropsychological measures, including Control and Sensitivity, which can be found in Figures S3 and S4. The brain plots in Figures~\ref{fig:brainplot}, S3, and S4 were created using BrainNet by inputting the Shen-268 atlas corresponding surface and volume files. Using the ROI drawing function in BrainNet, we visualized the covariance estimates of all fMRI conditions across the whole brain.  The circle plot (see Figure \ref{fig:circle plot}) is drawn on BioImage Suite Web developed by Yale \textcite{papademetris2006bioimage}; the circle plot showed the top 5 positive and top 5 negative functional connectivity edges associated with Sociability. We also did a corresponding brain plot for each circle plot, showing the position of the top 20 brain nodes and their covariance estimates with the Sociability category.

Brain connectome under the Eyes condition showed strongest associations with Sociability. In Figure~\ref{fig:brainplot}, a lighter color suggested that this brain region had stronger links with Sociability measures. The light yellow and light blue meant the brain region was positively or negatively associated with behavior variables within the Sociability category correspondingly. Across fMRI conditions, we found the rest condition showed generally lower association with Sociability than task conditions. This observation suggested that task-based connectomes contained more information about Sociability.  The brain plot for the Average condition showed that its association with Sociability fell between those of the rest and task conditions. Among the four task conditions, Eyes condition has the strongest link with Sociability.

To answer this question whether the Eyes condition consistently show strong associations with a range neuropsychological measures, we analyzed Control (Figure S3) and Sensitivity (Figure S4). Control was selected for its focus on cognitive behaviors, often contrasted with the Sociability dimension. Figure S3 showed that the Eyes condition had the highest covariance estimates. The SST condition, with higher covariance estimates for Control than Sociability, remained lower than Eyes condition. We examined Sensitivity due to its unique biomarker profile in Figure~\ref{fig:spiderplot}. Figure S4 revealed that strong links exist between the brain and Sensitivity under both Eyes and gradCPT conditions. These results confirmed that brain connectivity had the strongest associations with multiple neuropsychological measures under the Eyes condition including Sociability, Sensitivity, and Control. GradCPT showed a notable association with Sociability and Sensitvity, but a weak association with Control.  Together, our results show that certain tasks may be more optimal for specific outcomes than others.

Figure~\ref{fig:brainplot} and S2 to compare differences of the functional biomarkers across fMRI conditions.  Figure S2 used different colors to classify different functional regions in brain plots.  By matching the location of lighter brain regions in Figure~\ref{fig:brainplot} with the functional regions in Figure S2, we notice that under many fMRI conditions, a large proportion of brain regions with lighter colors (indicating stronger association with Sociability) are located in fronto-parietal areas. This observation aligned with Figure~\ref{fig:spider_task}, demonstrating the significance of fronto-parietal biomarkers for the Sociability category. Despite this general pattern, different fMRI conditions exhibited unique features. The Eyes condition had high connectivity nodes in the fronto-parietal, motor, and limbic regions. For gradCPT, EN-back, and SST, high covariance nodes were mainly in fronto-parietal areas. Rest condition showed high covariance in default mode and fronto-parietal regions, while the Average condition had large covariance nodes across default mode, fronto-parietal, and motor regions.

Figure~\ref{fig:circle plot} revealed distinct patterns of brain connectivity associated with Sociability across various fMRI conditions, highlighting the complex and task-dependent nature of social cognition in the brain.
This figure displayed connectivity among top Sociability-associated brain nodes. Red lines showed connections between top 5 positive nodes, blue lines for top 5 negative nodes.  For the rest condition, we observed positive connections (red) in frontal and temporal regions, while negative connections (blue) in occipital and cerebellar areas. GradCPT exhibited strong frontal and temporal connectivity, consistent with its role in attention and cognitive control. EN-back displayed widespread connectivity across multiple brain regions, reflecting the complex cognitive processes involved in this working memory task. The Eyes condition showed distinct connectivity patterns, with strong positive connections among prefrontal, limbic, and motor regions. This unique pattern might be related to the social and emotional processing demands of this task. SST demonstrated notable prefrontal-cerebellar positive connectivity and parietal-temporal negative connectivity, possibly reflecting the inhibitory control processes central to this task.

\section{Discussion}

In this study, we aim to address whether and how we can improve the cost-effectiveness of fMRI-based neuroimaging studies by using tailored fMRI tasks during data collection to investigate a range of neuropsychological measures. With generally satisfactory predictive power obtained by fitting our Bayesian generative network model for connectome-based predictive modeling, we have a robust foundation to investigate differences in predictive power and associated neuroimaging biomarkers between resting and various task conditions.  Notably, we found a lower prediction accuracy averaging across fMRI conditions then other fMRI conditions. This reduction in accuracy may be attributed to the averaging effect, which minimized the unique brain circuit perturbations from well-designed fMRI tasks related to different neuropsychological categories. In addition, we found limited prediction accuracy using the EN-back condition to model certain negative emotions.  This finding highlights that specific fMRI conditions may be more appropriate for predicting certain neuropsychological categories than others; there is a unique pairing of fMRI tasks and neuropsychological measures that should not be ignored when designing well-powered fMRI studies. We compared optimal fMRI conditions for each of the eight neuropsychological categories included in the study: BriefSymp, Control, Distress, Empathy, NegEmo, PosEmo, Sensitivity, and Sociability. This insight is essential for other researchers aiming to achieve higher power when studying specific outcomes with limited resources.

When comparing functional fingerprints associated with different neuropsychological outcomes, our study revealed that neuropsychological categories shared common functional biomarkers but maintained distinct differences. For instance, the fronto-parietal and motor functional systems were found to be associated with a range of neuropsychological outcomes with exceptions. The motor functional system was not found to contribute to Sensitivity outcomes. These unique biomarker profiles offered insights into the distinct characteristics of each neuropsychological category, suggesting avenues for future research. Investigating these unique biomarkers could help identify objective diagnostic features for psychiatric disorders.

In addition to comparing neuropsychological categories, we also conducted a comparable analysis of the neuroimaging biomarkers under different fMRI conditions. We found that the Eyes condition had the most medial-frontal biomarkers across task conditions. EN-back showed smaller involvement in the default mode network than others. Interestingly, we observed that the Eyes condition showed strong brain-behavior links with a range of behaviors including control, sensitivity, and sociality. GradCPT demonstrated strong links with Sensitivity and Sociability but weaker links with Control. The Eyes condition was found with strong predictive power for a range of outcomes.

Our study has limitations. The classification of behavioral categories may not have been sufficiently accurate, and there may have been inherent differences among variables in their psychometric properties within each neuropsychological category. Future studies should focus on refining behavioral classifications and exploring within-category variations to enhance the precision of predictions and biomarker identification.


\clearpage

\printbibliography

\end{document}


\maketitle

\section{Additional Figures}

\begin{figure}[h]
  \centering
  \includegraphics[width=0.9\textwidth]{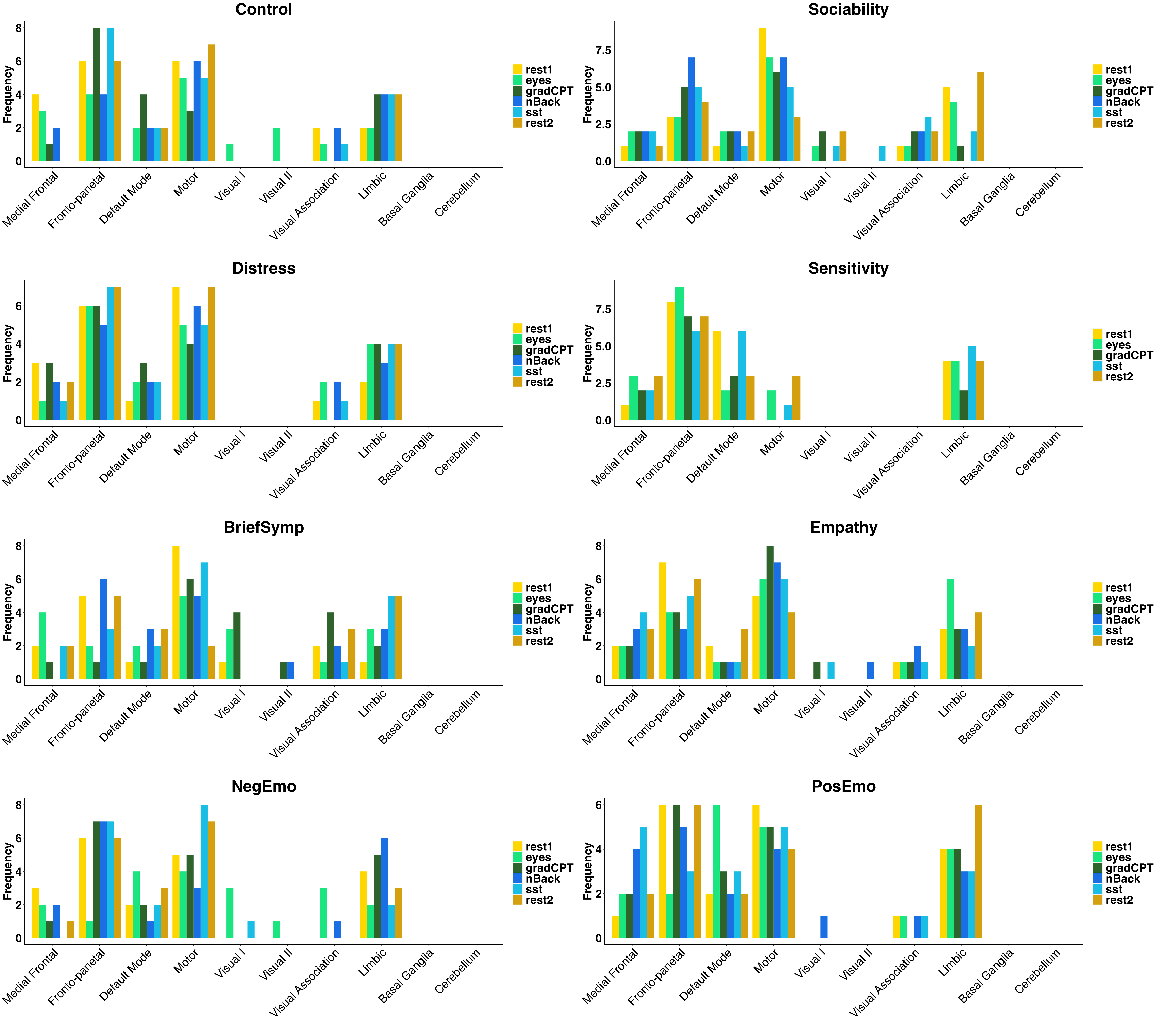}
    \caption {Barplot of Functional Labels by Outcome Category for each task condition}
  \label{fig:S1}
\end{figure}

\begin{figure}[h]
  \centering
  \includegraphics[width=0.9\textwidth]{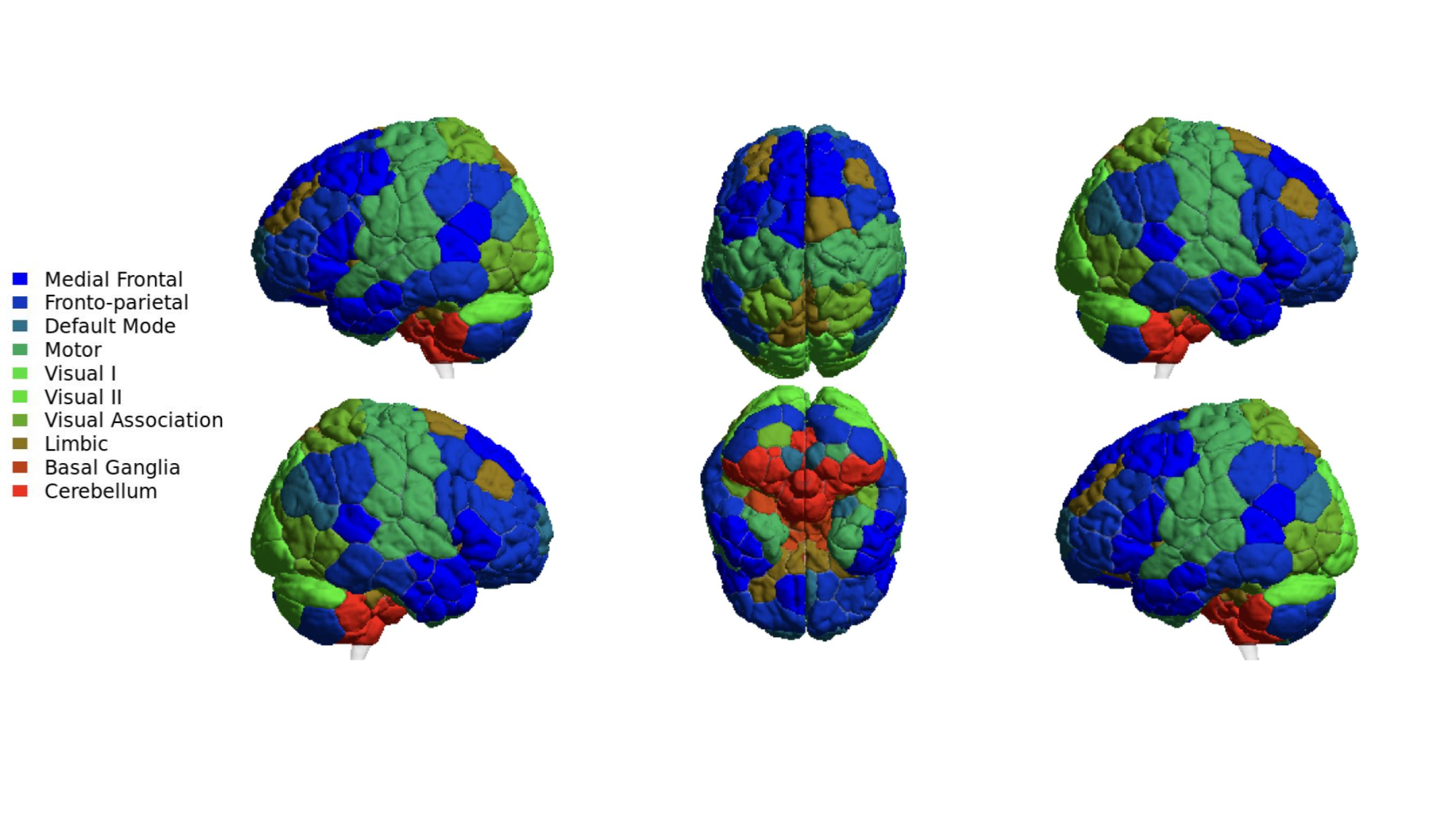}
    \caption {Brain plot for Anatomical Regions}
  \label{fig:anatomical}
\end{figure}

\begin{figure}[h]
  \centering
  \includegraphics[width=1\textwidth]{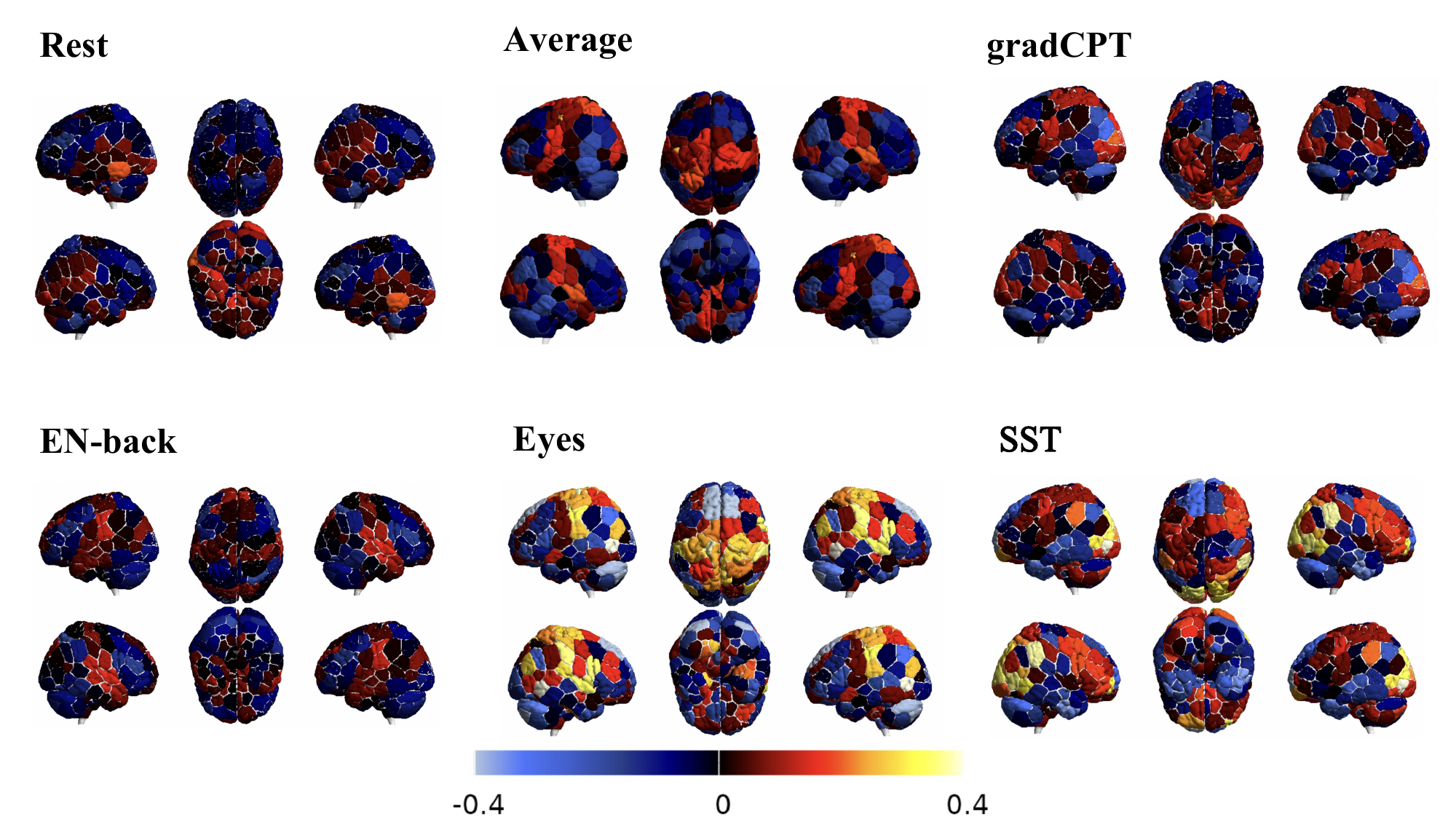}
    \caption {\textbf{ Covariance Estimates Between Brain Regions and Control Across Various fMRI Conditions}This figure visualizes the covariance estimates of 268 brain nodes across different fMRI conditions: Rest, Average, gradCPT, EN-back, Eyes, and SST. The color scale indicates the magnitude of covariance, ranging from -0.4 (blue) to 0.4 (yellow). With lighter color indicates higher absolute value of covariance}
  \label{fig:control_cov}
\end{figure}

\begin{figure}[h]
  \centering
  \includegraphics[width=1\textwidth]{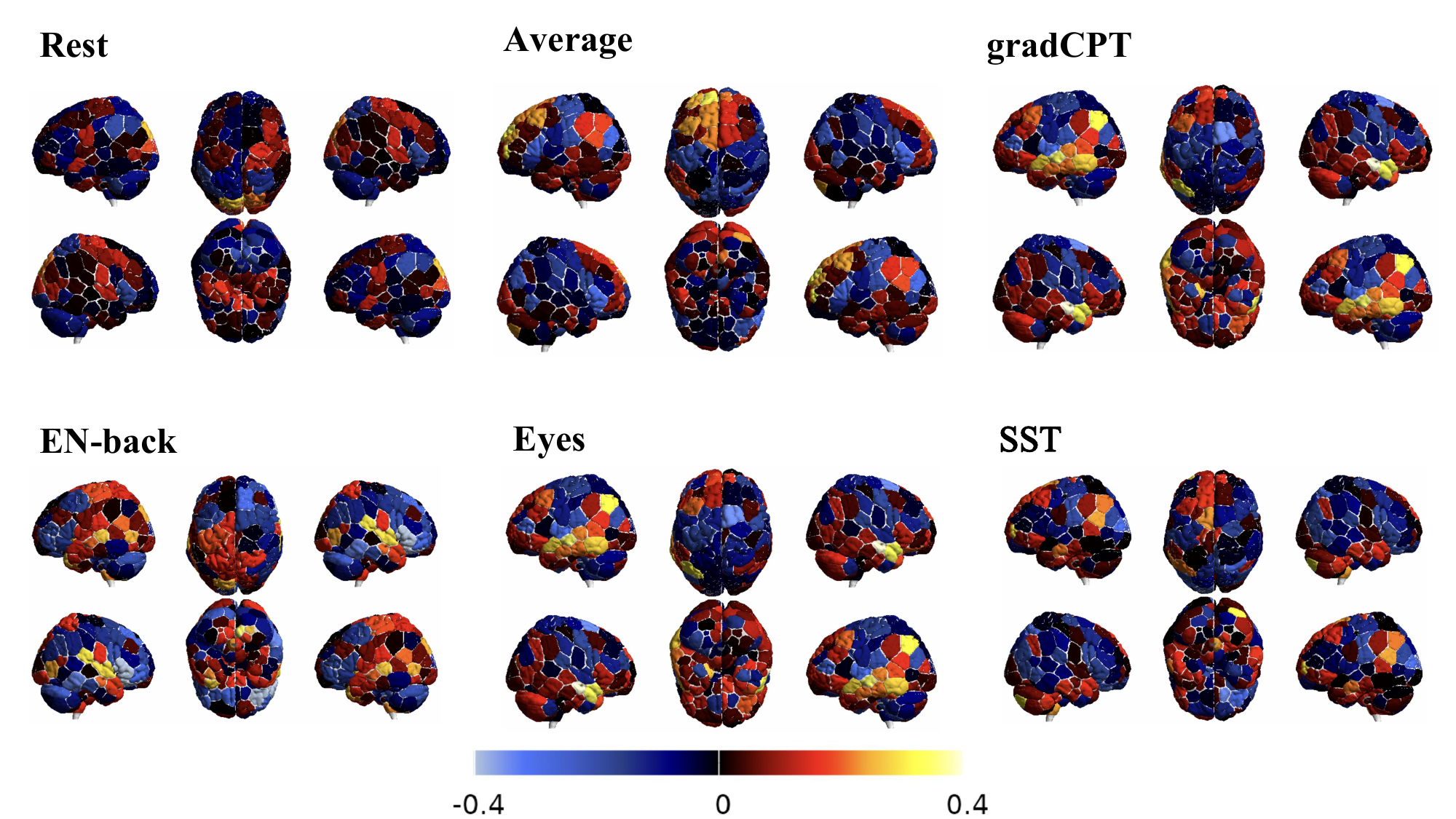}
    \caption {\textbf{ Covariance Estimates Between Brain Regions and Sensitivity Across Various fMRI Conditions}This figure visualizes the covariance estimates of 268 brain nodes across different fMRI conditions: Rest, Average, gradCPT, EN-back, Eyes, and SST. The color scale indicates the magnitude of covariance, ranging from -0.4 (blue) to 0.4 (yellow). With lighter color indicates higher absolute value of covariance}
  \label{fig:sensitivity_cov}
\end{figure}

\begin{figure}[h]
  \centering
  \includegraphics[width=0.8\textwidth]{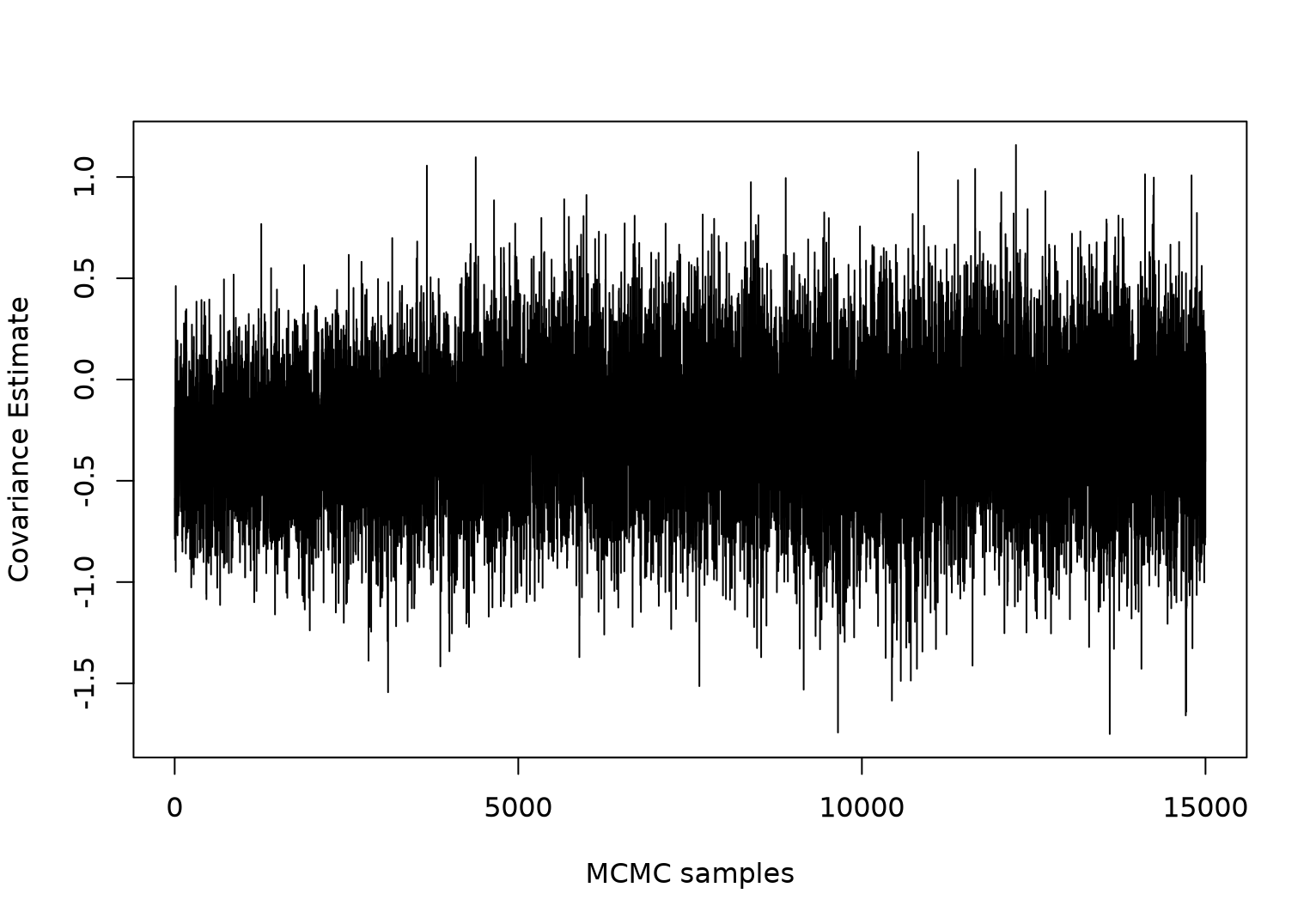}
    \caption {\textbf{Traceplot: covariance estimate between anxiety and a brain node under gradCPT task} The x-axis represents 15000 MCMC samples, and the y-axis represents the covariance estimate between anxiety variable and a brain node under gradCPT task. }
  \label{fig:traceplot}
\end{figure}

\clearpage

\section{Additional Tables}
This section can include tables that are supplementary to the main text.

\begin{table}[h]
\centering
\begin{tabular}{l r}
\toprule
\textbf{Diagnosis} & \textbf{Number of Patients} \\
\midrule
Depression & 36 \\
Anxiety disorders & 21 \\
Attention Deficit Hyperactivity Disorder (ADHD) & 16 \\
Bipolar Disorder & 14 \\
Schizophrenia & 10 \\
Borderline personality disorder & 6 \\
Panic disorder & 5 \\
Post-Traumatic Stress Disorder (PTSD) & 3 \\
Obsessive-Compulsive Disorder (OCD) & 1 \\
Antisocial personality disorder & 1 \\
Avoidant personality disorder & 1 \\
Anorexia nervosa & 1 \\
\bottomrule
\end{tabular}
\caption{Summary of Patient Diagnoses}
\label{tab:diagnosis_summary}
\end{table}

\begin{table}[ht]
\centering
\resizebox{1\textwidth}{!}{
\begin{tabular}{lcccccccc}
\toprule
\textbf{Variable} & \textbf{BriefSymp} & \textbf{NegEmo} & \textbf{PosEmo} & \textbf{Empathy} & \textbf{Distress} & \textbf{Sociability} & \textbf{Control} & \textbf{Sensitivity} \\
\midrule
Intercept & $0.458^{***}$ & $0.467^{***}$ & $0.449^{***}$ & $0.410^{***}$ & $0.490^{***}$ & $0.440^{***}$ & $0.428^{***}$ & $0.490^{***}$ \\
 & $(0.021)$ & $(0.026)$ & $(0.027)$ & $(0.021)$ & $(0.034)$ & $(0.028)$ & $(0.036)$ & $(0.033)$ \\
Rest2 & $-0.031$ & $-0.050$ & $-0.020$ & $0.047$ & $-0.084^{.}$ & $0.055$ & $-0.008$ & $-0.075$ \\
 & $(0.029)$ & $(0.037)$ & $(0.039)$ & $(0.030)$ & $(0.048)$ & $(0.040)$ & $(0.051)$ & $(0.047)$ \\
Average & $-0.031$ & $-0.035$ & $-0.043$ & $0.011$ & $-0.112^{*}$ & $-0.038$ & $-0.025$ & $-0.100^{.}$ \\
 & $(0.029)$ & $(0.037)$ & $(0.039)$ & $(0.030)$ & $(0.048)$ & $(0.040)$ & $(0.051)$ & $(0.047)$ \\
EN-back & $0.016$ & $-0.190^{***}$ & $-0.077^{.}$ & $0.049$ & $-0.016$ & $0.003$ & $0.073$ & $0.026$ \\
 & $(0.029)$ & $(0.037)$ & $(0.039)$ & $(0.030)$ & $(0.048)$ & $(0.040)$ & $(0.051)$ & $(0.047)$ \\
SST & $-0.003$ & $-0.010$ & $0.002$ & $0.037$ & $-0.067$ & $-0.016$ & $0.055$ & $-0.026$ \\
 & $(0.029)$ & $(0.037)$ & $(0.039)$ & $(0.030)$ & $(0.048)$ & $(0.040)$ & $(0.051)$ & $(0.047)$ \\
Eyes & $-0.029$ & $0.012$ & $0.016$ & $0.075^{*}$ & $-0.078$ & $0.008$ & $0.069$ & $-0.070$ \\
 & $(0.029)$ & $(0.037)$ & $(0.039)$ & $(0.030)$ & $(0.048)$ & $(0.040)$ & $(0.051)$ & $(0.047)$ \\
gradCPT & $0.017$ & $-0.013$ & $-0.036$ & $0.053^{.}$ & $0.012$ & $0.062$ & $0.025$ & $-0.035$ \\
 & $(0.029)$ & $(0.037)$ & $(0.039)$ & $(0.030)$ & $(0.048)$ & $(0.040)$ & $(0.051)$ & $(0.047)$ \\
\midrule
\textbf{R-squared} & $0.106$ & $0.493^{***}$ & $0.192$ & $0.297$ & $0.362$ & $0.408$ & $0.340$ & $0.433$ \\
\bottomrule
\end{tabular}
}
\textit{}\\Standard errors in parentheses \\ Signif. codes: 0 '***' 0.001 '**' 0.01 '*' 0.05 '.' 0.1 ' ' 1
\smallskip
\begin{flushleft}
Note. BriefSymp: Psychological Symptoms Inventory, NegEmo: Negative Emotional Spectrum, PosEmo: Positive Emotional Spectrum, Empathy: Empathy Engagement Scale, Distress: Emotional Distress Spectrum, Sociability: Sociability and Vitality Scale, Control: Self-Regulation Control Measures, Sensitivity: Sensory and Emotional Awareness Scale. 
\end{flushleft}
\caption{\textbf{Regression of Prediction Accuracy on Task Labels for Each category}}
\label{tab:regression_prediction_tasklabels_maineffect}
\end{table}

\begin{table}[ht]
\centering
\begin{tabular}{lcccc}
\toprule
\textbf{Coefficient} & \textbf{Estimate} & \textbf{Std. Error} & \textbf{t value} & \textbf{Pr($>|t|$)} \\
\midrule
(Intercept)                    & -0.030871 & 0.003127 & -9.872 & $< 2\text{e}-16$ *** \\
Medial Frontal & -0.005684 & 0.004065 & -1.398 & 0.162 \\
Fronto-parietal & 0.065978 & 0.003941 & 16.741 & $< 2\text{e}-16$ *** \\
Motor          & 0.029553 & 0.003700 & 7.987 & $1.40\text{e}-15$ *** \\
Visual I       & 0.005119 & 0.004544 & 1.127 & 0.260 \\
Visual II      & -0.023928 & 0.005613 & -4.263 & $2.02\text{e}-05$ *** \\
Visual Association & 0.035722 & 0.004544 & 7.862 & $3.84\text{e}-15$ *** \\
Limbic         & 0.067634 & 0.004037 & 16.753 & $< 2\text{e}-16$ *** \\
Basal Ganglia  & 0.017629 & 0.004065 & 4.337 & $1.45\text{e}-05$ *** \\
Cerebellum     & 0.051504 & 0.004011 & 12.841 & $< 2\text{e}-16$ *** \\
\bottomrule
\end{tabular}
\caption{Summary of Regression Analysis of Covariance Estimate}
\label{tab:regression_cov_functionallabels_maineffect}
\end{table}

\begin{table}[ht]
\centering
\resizebox{1\textwidth}{!}{
\begin{tabular}{lcccccccc}
\toprule
\textbf{Variable} & \textbf{BriefSymp} & \textbf{NegEmo} & \textbf{PosEmo} & \textbf{Empathy} & \textbf{Distress} & \textbf{Sociability} & \textbf{Control} & \textbf{Sensitivity} \\
\midrule
Intercept & -0.007  & $-0.045^{***}$ & $-0.061^{***}$ & -0.015 & $-0.033^{***}$ & $0.022.$ & $-0.078$ & $-0.034^{*}$\\
& {(0.004)}  & {(0.007)} & {(0.007)} & {(0.011)} & {(0.010)} & {(0.012)} & {(0.014)} & {(0.014)} \\
Medial Frontal & $0.011.$  & 0.007 & 0.005 & $-0.058^{***}$ & 0.005 & $-0.013$ & $-0.013$ & $-0.037^{*}$ \\
& {(0.006)}  & {(0.009)} & {(0.009)} & {(0.014)} & {(0.013)} & {(0.016)} & {(0.018)} &  {(0.018)} \\
Fronto-parietal& $0.049^{***}$  & $0.086^{***}$ & $0.072^{***}$ & $0.086^{***}$ & $0.070$ & 0.015 & $0.089^{***}$ & $0.057$ \\
& {(0.006)}  & {(0.009)} & {(0.009)} & {(0.014)} & {(0.012)} & {(0.015)} & {(0.018)} & {(0.017)} \\
Motor   & $-0.038^{***}$  & $0.014^{***}$ & $0.157^{***}$ & $-0.023.$ & $0.006$ & $-0.005$ & $0.123^{***}$ &  $0.063^{***}$\\
&  {(0.006)} & {(0.009)} & {(0.009)} & {(0.013)} & {(0.012)} & {(0.014)} & {(0.016)} & {(0.016)} \\
Visual I & $-0.036^{***}$  & $0.027^{*}$ & $0.095^{***}$ & $-0.056^{***}$ & $-0.016$ & $-0.081^{***}$ & $0.096^{***}$ &  $0.007$\\
& {(0.007)}  & {(0.011)} & {(0.010)} & {(0.016)} & {(0.014)} & {(0.017)} & {(0.020)} & {(0.020)} \\
Visual II & -0.002  & 0.009 & $-0.059^{***}$ & $-0.014$ & $0.002$ & $-0.109^{***}$ & -0.011 & $-0.074^{**}$ \\
& {(0.009)}  & {(0.013)} & {(0.012)} & {(0.020)} & {(0.018)} & {(0.022)} & {(0.025)} & {(0.024)} \\
Visual Association&  -0.006 & $0.036^{***}$ & $0.079^{***}$ & $0.021$ & $0.053^{***}$ & $-0.084^{***}$ & $0.151^{***}$ & $0.077^{***}$ \\
& {(0.007)} & {(0.011} & {(0.010)} & {(0.016)} & {(0.014)} & {(0.017)} & {(0.020)} & {(0.020)} \\
Limbic &  $0.015^{*}$ & $0.092^{***}$ & $0.098^{***}$ & $0.068^{***}$ & $0.053^{***}$ & 0.023 & $0.154^{***}$ & $0.089^{***}$ \\
& {(0.006)}  & {(0.009)} & {(0.009)} & {(0.015)} & {(0.013)} & {(0.016)} & {(0.018)} & {(0.018)}\\
Basal Ganglia& $-0.022^{***}$  & $0.036^{***}$ & $0.044^{***}$ & $0.009$ & $0.023.$ & -0.016 & $0.076^{***}$ & $0.025$ \\
& {(0.006)}  & {(0.009)} & {(0.009)} & {(0.015)} & {(0.013)} & {(0.016)} & {(0.018)} & {(0.018)} \\
Cerebellum&  $0.050^{***}$ & $0.100^{***}$ & $0.025^{**}$ & $0.060^{***}$ & $0.083^{***}$ & $-0.063^{***}$ & $0.069^{***}$ & $0.038^{*}$ \\
& {(0.006)}  & {(0.009)} & {(0.008)} & {(0.014)} & {(0.013)} & {(0.015)} & {(0.018)} & {(0.017)} \\
\midrule
\textbf{R-squared} & $0.033^{***}$  & $0.025^{***}$ & $0.077^{***}$ & $0.032^{***}$ & $0.017^{***}$ & $0.021^{***}$ & $0.034^{***}$ & $0.021^{***}$ \\
\bottomrule
\end{tabular}
}
\textit{}\\Standard errors in parentheses \\ Signif. codes: 0 '***' 0.001 '**' 0.01 '*' 0.05 '.' 0.1 ' ' 1
\smallskip
\begin{flushleft}
Note. BriefSymp: Psychological Symptoms Inventory, NegEmo: Negative Emotional Spectrum, PosEmo: Positive Emotional,Empathy: Empathy Engagement Scale, Distress: Emotional Distress Spectrum, Sociability: Sociability and Vitality Scale, Control: Self-Regulation Control Measures, Sensitivity: Sensory and Emotional Awareness Scale
\end{flushleft}
\caption{Regression of Covariance Estimates on Functional Labels for Each Category}
\label{tab:regression_functionallabels}
\end{table}

\begin{table}[ht]
\centering

\resizebox{1\textwidth}{!}{
\begin{tabular}{lcccccccc}
\toprule
\textbf{Variable} & \textbf{BriefSymp} & \textbf{NegEmo} & \textbf{PosEmo} & \textbf{Empathy} & \textbf{Distress} & \textbf{Sociability} & \textbf{Control} & \textbf{Sensitivity} \\
\midrule
Intercept &  -0.001 & -0.004 & $0.019^{***}$ & 0.001 & -0.010 & -0.009 & 0.002 & 0.004 \\
& {(0.004)} & {(0.005)} & {(0.005)} & {(0.008)} & {(0.007)} & {(0.009)} & {(0.010)} & {(0.010)} \\
Eyes & -0.003 & 0.009 & -0.006 & -0.001 & 0.013 &  0.011& 0.009 &  0.003 \\
& {(0.005)} & {(0.007)} & {(0.008)} & {(0.012)} & {(0.010)} & {(0.012)} & {(0.015)} & {(0.014)} \\
GradCPT& $-0.012^*$ & $-0.015^{**}$ & $-0.015^*$ & $-0.020.$  & -0.010 & 0.000 & -0.009 & -0.021 \\
& {(0.005)} & {(0.007)} & {(0.008)} & {(0.012)} & {(0.010)} & {(0.012)} & {(0.015)} & {(0.014)} \\
N-Back  & $-0.011^*$ & -0.007& -0.009& -0.008& 0.003 & $0.022.$ & 0.004 & -0.008 \\
            & ${(0.005)}$ &{(0.007)} &{(0.008)} &{(0.012)} &{(0.010)} & {(0.012)} & {(0.015)} & {(0.014)} \\
Stop Signal& 0.008 & $0.023^{***}$ & $-0.025^{***}$ & $0.010$ & $0.028^{**}$ & 0.000 & 0.000 & 0.013 \\
& {(0.005)} & {(0.007)} & {(0.008)} & {(0.012)} & {(0.010)} & {(0.012)} & {(0.015)} & {(0.014)} \\
Second Resting & -0.005 &0.013  & -0.012 & -0.001 & 0.011 & 0.002 &  0.025. & 0.002 \\
& {(0.005)} & {(0.007)}  & {(0.008)} & {(0.012)} & {(0.010)} & {(0.012)} & {(0.015)} & {(0.014)} \\
Average&  -0.005 & 0.007 &  $-0.024^{**}$  & -0.001 & 0.009 & 0.006 & -0.007 & -0.010 \\
& {(0.005)} & {(0.007)} & {(0.008)} & {(0.012)} & {(0.010)} & {(0.012)} & {(0.015)} & {(0.014)} \\
\midrule
\textbf{R-squared} & $0.001^{***}$ & $0.002^{***}$ & $0.001^*$ & 0.001 & $0.002^*$  & 0.001 & 0.001 & 0.001 \\
\bottomrule
\end{tabular}
}
\textit{}\\Standard errors in parentheses \\ Signif. codes: 0 '***' 0.001 '**' 0.01 '*' 0.05 '.' 0.1 ' ' 1
\smallskip
\begin{flushleft}
Note. BriefSymp: Psychological Symptoms Inventory, NegEmo: Negative Emotional Spectrum, PosEmo: Positive Emotional, Empathy: Empathy Engagement Scale, Distress: Emotional Distress Spectrum, Sociability: Sociability and Vitality Scale, Control: Self-Regulation Control Measures, Sensitivity: Sensory and Emotional Awareness Scale
\end{flushleft}
\caption{Regression of Covariance Estimates on Task Labels for Each Category}
\label{tab:regression_tasklabels}
\end{table}

\begin{table}[H]
\centering
\caption{Number of Biomarkers across different conditions}
\label{table:number}
\begin{tabular}{lS[table-format=1.3]S[table-format=1.3]S[table-format=1.3]S[table-format=1.3]S[table-format=1.3]}
\toprule
Conditions & {Medial Frontal} & {Fronto-parietal} & {Default Mode} & {Motor} & {Visual I} \\
\midrule
Rest & 1.938 & 5.875 & 2.063 & 5.188 & 0.188 \\
Eyes & 2.375 & 3.875 & 2.625 & 4.875 & 1.000 \\
gradCPT & 1.750 & 5.500 & 2.375 & 4.625 & 0.875 \\
EN-Back & 2.143 & 5.286 & 1.857 & 5.429 & 0.143 \\
SST & 2.000 & 5.500 & 2.375 & 5.250 & 0.375 \\
Average & 2.000 & 5.375 & 2.500 & 4.375 & 0.875 \\
\midrule
\end{tabular}

\vspace{0.5cm}

\begin{tabular}{lS[table-format=1.3]S[table-format=1.3]S[table-format=1.3]S[table-format=1.3]S[table-format=1.3]}
\toprule
Conditions &  {Visual II} & {Visual Association} & {Limbic} & {Basal Ganglia} & {Cerebellum} \\
\midrule
Rest & 0.000 & 0.813 & 3.813 & 0 & 0 \\
Eyes & 0.375 & 1.250 & 3.625 & 0 & 0 \\
gradCPT & 0.125 & 0.875 & 3.125 & 0 & 0 \\
EN-Back & 0.286 & 1.714 & 3.143 & 0 & 0 \\
SST & 0.125 & 1.000 & 3.375 & 0 & 0 \\
Average & 0.000 & 0.375 & 3.875 & 0 & 0 \\
\midrule
\end{tabular}
\begin{flushleft}
Note. Each row represented the averaged number of top 20 functional biomarkers associated with each function system over eight behavioral categories for each fMRI condition. \sw{move to supplementary materials.}
\end{flushleft}
\end{table}